\documentclass[times,10pt,onecolumn]{article}

\usepackage{epsf}\epsfverbosetrue
\usepackage{graphics,epsfig,color}
\usepackage{graphicx,epsfig} 
\usepackage{alltt}
\usepackage{subfigure}
\usepackage{float}
\usepackage{url}
\usepackage{graphicx}
\usepackage{verbatim} 
\usepackage{footnote} 
\usepackage{latexsym} 
\usepackage{amsmath}
\usepackage{sidecap} 
\usepackage{color}
\usepackage{algorithm}
\usepackage{algpseudocode}
\usepackage{ioa_code}

\begin{document}

\title {Decoupling data dissemination from the mobile sink's trajectory in wireless sensor networks: Current Research and Open Issues \newline  \newline \newline \emph { \normalsize Report Submitted for the partial fulfilment of the Degree of Masters by Research (M2R)  \normalsize in Networks and Telecommunications \normalsize from  \normalsize Laboratory of Signals and Systems (L2S), \normalsize Sup$\acute{e}$lec,  \normalsize University of Paris Sud XI} \newline \newline \newline   \centering Stage Director: Dr. Aline Carneiro Viana\thanks{The author would like to thank Artur Ziviani from LNCC, Brazil for the helpful discussions and suggestions,
and for their very careful reading of the manuscript.}  \newline \newline   \centering Responsible of Masters: Dr. Pierre Duhamel \newline \newline \newline}
\author{\centering Submitted By : \Large  Mubashir Husain Rehmani}

\date{Version 1 \vspace{35pt} \\ June 27, 2008}



\maketitle
\pagebreak

\begin{abstract}
The aim of Wireless Sensor Networks (WSNS) is to accumulate data and in this perspective,
firstly the need of distributed data management emerges and gain vital importance. Secondly data
dissemination of this accumulated data has its own importance. Now distributed data management and
data dissemination is an essential paradigm in wireless sensor networks resulting in minimizing the
number of transmissions, eliminating the redundant data, conserve the energy, and thus resulting in
the overall increase in the lifetime of the network. Traditionally data dissemination has been done
by utilizing static sinks. These static sinks were not only prone to hotspots problem but also
decrease throughput, increase the number of transmitted packets, less energy conservation of sensor
nodes, and above all decrease the overall network lifetime. To deal with these aforementioned
problems the need of incorporating mobile sink arises. Now the paradigm of distributed data
management and data dissemination is shifting from static sink to mobile sink and now more and more
research work has been done in the domain of mobile sink wireless sensor networks.

In this report, firstly, we presents state of the art survey on Data Management and Data
Dissemination techniques with Mobile Sink. Moreover we classify these techniques into two ample
sub-categories. Under this classification, we identify, review, compare, and highlight these
techniques and their pros and cons. We do a SWOT (Strength, Weaknesses, Opportunities, Threats)
analysis of each scheme. We also discuss where each scheme is appropriate.

Secondly, we presents a new \emph{distributed data management} scheme which is based upon Random
Walk Based Membership Service to facilitate Data Dissemination in Mobile Sink based Wireless Sensor
Networks. Our proposed scheme efficiently deals with the aforementioned problems and we also
compare the characteristics of our proposed scheme with the state-of-the-art data-dissemination
schemes. We propose using Random Walks~(RWs) with uniformly distributed views to disseminate
data through the WSN with a controlled overhead. This is performed by the use of a Random Walk
Based Membership Service - the RaWMS. Our proposal solves then the problems generated when {\it
(a)} all nodes are storage motes, being no aggregation performed {\it (b)} one center node plays
the role of storage mote and aggregates data from all the other nodes {\it (c)} replication is
performed on all nodes in the network.

To the best of our knowledge, we are the first to propose an efficient data dissemination approach
(in terms of overhead, adaptiveness and representativeness) to allow a mobile sink to gather a
representative view of the monitored region covered by $n$ sensor nodes by only visiting {\it any} $m$ nodes, where hopefully $m << n$. 

\end{abstract}
\pagebreak
\tableofcontents
\pagebreak
\listoffigures
\listoftables
\pagebreak


\section{Introduction}
\label{sec:introduction}


%
%

With the advent and emergence of technology, Wireless Sensor Nodes (WSN) become more and more
abundant resulting in the creation of highly dense Wireless Sensor Networks (WSNs) containing
hundred to thousands of Wireless Sensor Nodes. This is due to the fact that day by day the
technology is getting cheaper and the fascinating applications of WSNs attracted attention of many
researchers and scientists for massive deployment of these types of networks.

The application of WSNs ranging from environmental monitoring like wildlife tracking, monitoring
volcanoes, habitat monitoring, forest fire detection, mine safety monitoring to military
applications like target detection, classification and tracking \cite{IEEEhowto:anis}, Sensor
Network based counter sniper system \cite{IEEEhowto:gyula} etc. The authors of \cite{IEEEhowto:i}
has indicated many applications areas in the context of multimedia enabled WSNs like industrial
process control, person locater information, advance health care delivery.

Provided all these advantages, the researchers come across many challenges to collect monitored
data in these kind of networks like connectivity, security,  data dissemination and collection,
consuming power and bandwidth, just to name a few \cite{IEEEhowto:david}.

A WSNs is composed of tiny wireless sensor nodes, which can either be static, mobile or hybrid of
two, depending upon the nature of application. These WSNs may be time-driven or event-driven
distributed systems and can be operated in unattended mode. These sensor nodes are scattered in a
random fashion and their main goal is to periodically sensed data, process it and send it to the
sink. Since these sensor nodes are energy constrained, we should take into account distributed data
management and data dissemination seriously, while accumulating, storing or disseminating sensed
data.



Since the main goal of WSNs is to collect data, so data management and data dissemination has vital
importance in this perspective. Here, data management means how to efficiently store collected data
so that it can be retrieved later. Otherwise, to guarantee their correct delivery and data
dissemination means data distribution in the network. Traditionally data dissemination has been
done by utilizing static sinks. These static sinks were not only prone to hotspots problem but also
decrease throughput, increase the number of transmitted packets, less energy conservation of sensor
nodes, and above all decrease the overall network lifetime. Some solutions to deal with hotspots
problem is: to distribute the workload from hotspots to those nodes who are in the vicinity of
these hotspots, to deploy multiple static sinks, or to make sink move \cite{IEEEhowto:yanzhong}.
To deal with these aforementioned problems the need of incorporating mobile sink arises.

Now the paradigm of distributed data management and data dissemination is shifting from static sink
to mobile sink and now more and more research work has been done in the domain of Mobile Sink
Wireless Sensor Network (MSWSN).

\cite{IEEEhowto:ioannis} has indicated many advantages of utilizing mobile sink. Their findings
demonstrates that with very limited sink mobility, the overall success rate can be improved by 50\%
and energy dissipation can be reduced to 30\%. We can achieve nearly 100\% success rates and
further reduce the energy consumption if we let the sink fully mobile.


Thus, the trajectory of the Mobile Sink has gained more and more attention and many solutions have
been proposed to optimize sink trajectory to accumulate data from sensor nodes.

To the best of our knowledge, we are the first who are investigating the state-of-the-art of data
management and data dissemination schemes with Mobile Sink Wireless Sensor Networks (MSWSN) and
this is the first survey of its kind in this domain. Our contribution in this report is to
understand existent schemes and classify them on the basis of mobile sink trajectory and
data-gathering mechanisms. Inspired by the work of Albert Humphery of Standford University
\cite{IEEEhowto:bg}, we have done a SWOT (Strength, Weaknesses, Opportunities, Threats) analysis of
each scheme. We also discuss where each schemes are appropriate. We conclude with possible future
research directions.

The remainder of this report is organized as follows: In Section~\ref{sec:data}, we discuss
the advantages of data collection with mobile sinks. We then classify the existent approaches in
Section~\ref{sec:classification}. We describe these approaches in detail in
Section~\ref{sec:proactive} and~\ref{sec:reactive}. Section~\ref{sec:ourproposal} is equip with
our contribution and proposal. In Section~\ref{sec:handling}, we analyze different parameters and
the trade-off between data dissemination and mobile sinks trajectory. We then discuss and mention
open issues in Section~\ref{sec:discussion} and finally, in Section~\ref{sec:conclusion}, we
conclude this report.

\section{Data collection with mobile sinks}
\label{sec:data}

Incorporating Mobile Sink to accumulate data in WSNs is advantageous and results in energy
conservation  of sensor nodes. Nevertheless, it will depend upon the employed technique. In
addition to that there is no need of full network connectivity to gather data from the network.
Moreover, it is not obligatory for the sensor nodes to find routes to sinks which generally causes
hotspots problem.

With mobile sink sparse and disconnected networks can be easily handled. On the one hand, we can
increase throughput and data fidelity, and on the other hand, we can reducing the overhead in
routing control. Security and robustness can be enhanced because multi-hop routing is not required,
mobile sink can navigate through or bypass problematic regions where sensor devices cannot
operate, and above all mobile sink increases the overall lifetime of the network where energy is a
scarce resource. Besides this, mobile sink avoids sensor to sink path maintenance in the network
and makes the network free to self-organize \cite{IEEEhowto:ioannis},\cite{IEEEhowto:aka}.

In this section we are elaborating some advantages of utilizing mobile sink:

\begin{itemize}
\item{Energy Conservation and Overall lifetime of Network:}
Sensor nodes are highly energy constraint and their batteries are not replenishable because in some
applications like behind the enemy lines or inhospitable terrains, physically reaching the nodes
and replace or recharge the batteries is no longer feasible.

Since energy is scarce resource in sensor nodes, the incorporation of Mobile Sink led to the
conservation of energy of sensor nodes by reducing the communication.

Besides this a mobile sink has the possibility to reach directly to sensor nodes (this will depend
upon the scheme used), so sensor devices can reduce their transmission range to the lowest value
required further decreasing the energy consumption.

Node duty state can be set according to sink visits thus conserve energy of sensor nodes.

Sensor nodes does not have to send packets to multiple hops or disseminate packets to the whole
network thus resulting in the reduction in number of transmitted packets which definitely decrease
the energy consumption of nodes and thus increase the overall lifetime of the  network.

\item{Network Connectivity:}
Through Mobile Sink, network is considered virtually fully connected in a sense that a Mobile Sink
can traverse the whole network or some specific regions in the network without each and every
sensor node being physically connected.

\item{Hotspots Problem:}
Schemes that uses multi-hop routing to forward data to particular nodes in the network results in
the drainage of battery of these particular nodes and causes hotspots problem. With mobile sink
hotspots problem can be minimized to a certain extent or sometimes, completely removed.

\item{Increased Throughput and Data Fidelity:}
Latency can be reduced because the sojourn state of Mobile Sink increases the throughput and thus
resulting in data fidelity. Mobile sinks also decrease the requirement for routes definition in the
network, and by consequence reduce the overhead in routing control. There will be less number
of communication in the network. This decreases the number of collision, increasing then the
throughput.

\item{Reduction in Probability of Transmission Errors and Collisions:}
There is a great influence of incorporating Mobile Sink on the reduction of probability of
transmission errors  and collisions. Because in the presence of mobile sink there is no need of
multi hop routing or we can say that the number of hops reduces and thus less number of
retransmission occurs.

\item{Operational Cost of the Network:}
A mobile sink allows monitoring a region with fewer sensor devices thus decreasing the operational
cost of the network. For instance, in problematic regions, we can decrease the number of nodes and
just place these nodes over certain locations from where we want to collect data. It can be
then supposed these particular nodes are not connected to the network (no need of intermediate
nodes to connect these nodes among them or to a base station). So in this case mobile sink can
visit to these particular nodes and thus to decrease the operational cost of the network.

Besides this sensor nodes creates small isolated networks and due to the presence of mobile sink
there  is no need to connect these small isolated networks, again decreasing the operational cost
of the network.

\item{Enhancement in Security:}
Since there is no need of multi-hop routing in the presence of Mobile Sink and data does not have
to traverse  multiple hop across potentially compromised nodes, security can then be enhanced and
an intruder cannot sniff the data easily. If an intruder is able to sniff the data packets then he
can only receive information regarding that area because the information is not disseminated in the
whole network.

\item{Accessibility to Problematic Regions:}
Mobile sink can access to problematic regions and can easily collect the data where human
accessibility is not feasible like inhospitable terrain, behind the enemy lines, etc.
\end{itemize}

%
%

\section{Classification}
\label{sec:classification}
%

%



\begin{table}[!t]
\renewcommand{\arraystretch}{2}
\caption{Proactive and Reactive Data-Dissemination and Data-Management Schemes}

\label{table_example23}
\centering
\begin{tabular}{p{5cm}||p{1.5cm}||p{1.5cm}}

\hline

\bfseries \footnotesize {Protocol Name} & \bfseries  \footnotesize Proactive & \bfseries \footnotesize Reactive \\
\hline\hline

\footnotesize Our Proposal & \footnotesize Yes & \footnotesize No\\
\hline
\footnotesize Moving Schemes for Mobile Sink in WSN & \footnotesize Yes & \footnotesize No\\
\hline
\footnotesize Coordinate Magnetic Routing for MSWSN & \footnotesize Yes & \footnotesize No\\
\hline
\footnotesize FLOW  & \footnotesize Yes & \footnotesize No\\
\hline
\footnotesize WEDAS & \footnotesize Yes & \footnotesize No\\
\hline
\footnotesize Locators of Mobile Sink for WSNs & \footnotesize No & \footnotesize Yes \\
\hline
\footnotesize Interest Dissemination with Directional Antennas for WSNs with Mobile Sinks & \footnotesize No & \footnotesize Yes\\
\hline
\footnotesize Data MULES & \footnotesize No & \footnotesize Yes\\
\hline
\footnotesize MobiRoute  & \footnotesize No & \footnotesize Yes\\
\hline
\footnotesize Efficient Data Propagation Strategies in WSNs using a Mobile Sink & No & Yes\\
\hline

\end{tabular}
\end{table}


While reviewing the state of the art, we encountered many approaches with distinguished
characteristics. For instance, some approaches proposed that the mobile sink should visit some
nodes while other proposed that mobile sink must visit each and every node in order to collect
data. If the mobile sink visits all nodes then no network organization in terms of data management
and data dissemination is performed and these are due to the fact that sensor nodes:

\begin{itemize}
\item become awake during all the retrieval period;
\item keep collected data in the memory until their sink retrieval, which can
restrict new data collection in the case of memory resource limitation;
\item nodes represent points of failures, since no replication is performed.\\
\end{itemize}

In fact, there is a need to remove this kind of visit of mobile sink. Therefore researchers are
paying their  attention to optimize mobile sink trajectory provided that the defined mobile sink
trajectory allow us to collect data of whole network. So in this report, we are classifying Data
Management and Data Dissemination Schemes with Mobile Sinks into two ample categories: Proactive
and Reactive.

Proactive means that the network establishes first an organization, in terms of \emph {data}
\emph{dissemination} and \emph{data storage}, deciding thus, which nodes are responsible for
keeping the data. The mobile sink can then visit these nodes and collect data.


In Reactive Data Management and Data Dissemination Schemes, no previous network organization in
terms of data management and data dissemination is performed. The data is gotten by the sink during
the period of visits. In spite of that, some techniques have proposed to optimize/manage the data
collection by the sink in order to avoid the visit of all nodes. The Mobile Sink will send queries
and in reply the Sensor Nodes will send data irrespective of whether the routes between the nodes
are established or not and the network is connected or not.

Table \ref{table_example23} summarizes that which protocol is proactive and which protocol is reactive in terms of Data-Dissemination and Data-Management.

\section{Proactive approaches}
\label{sec:proactive}



\begin{figure}[!t]
\centering
\setlength\fboxsep{0pt}
\setlength\fboxrule{0.5pt}
\includegraphics[height=60mm,width=120mm]{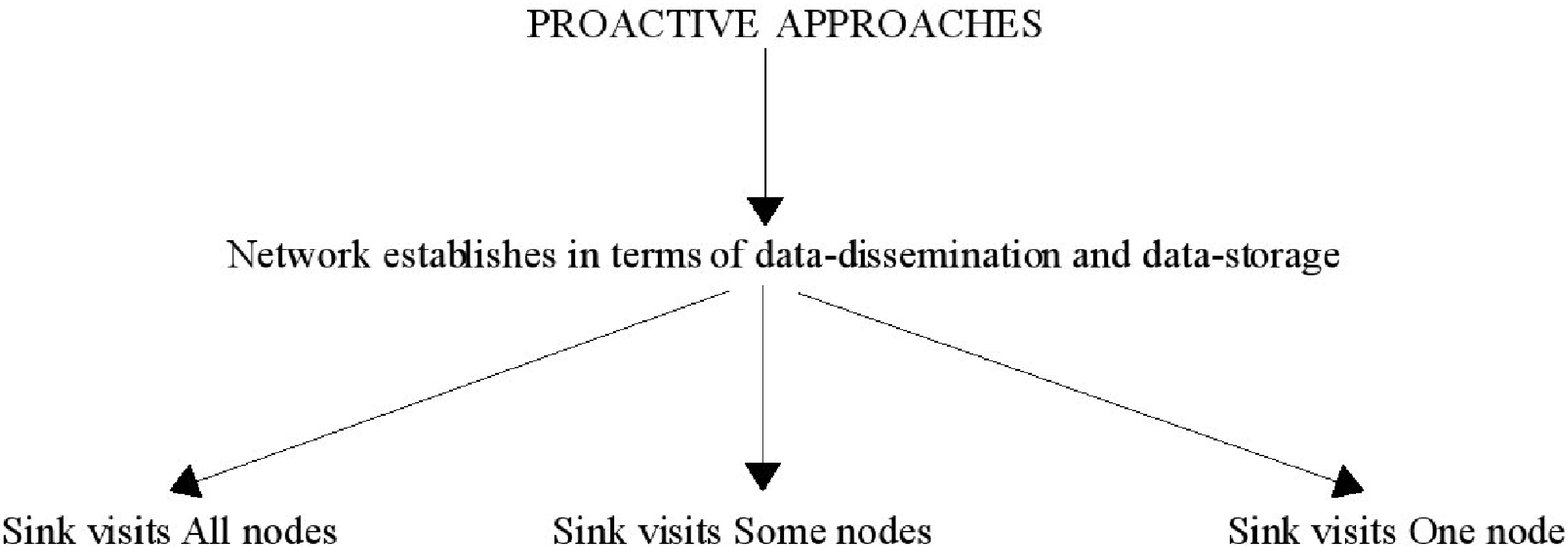}
\caption{Proactive schemes indicating the need of Mobile Sinks Visit} \label{fig_sim1}
\end{figure}

In Proactive data dissemination and data management schemes,
several mechanisms can be used for defining the mobile sink's trajectory. In fact, the more appropriate
mechanism will depend on the applied data dissemination approach. But still, one important issue to guarantee is
that the defined sink trajectory should allow it to collect data of the whole network. For instance a mobile sink
may visit all the nodes, visit some nodes, or visit one node as depicted in Fig. \ref{fig_sim1}.\\

One simple example of this kind of trajectory is the case when mobile sink visits all nodes, where
nodes  store their own data being no replication or aggregation performed, as in the case of \cite{IEEEhowto:habib}.


A simple example of mobile sink trajectory is the case when mobile sink visits some sensor node
(these are storage nodes that stores data of all other sensor nodes). This reduces the trajectory
length of mobile sinks but requires some organization to be performed in order to select the storage nodes. For instance, in \cite{IEEEhowto:shih}, the mobile sink's trajectory depends upon received data. It means that the mobile sink will follow a particular trajectory and visits some nodes. The authors of \cite{IEEEhowto:rahul} proposes fixed trajectory for mobile sink, which means that the mobile sink will visit some nodes, not the whole network.

Another example is that when mobile sink just visit only one sensor node in the network then it is
obligatory that this one node must collect data of all other remaining nodes.

The following sections describe some proactive schemes. Our goal here is not to provide a
exhaustive description of related works in the literature, but to describe their general idea.



\subsection{Our Proposal:}

We presents  \footnote {See Section 6 for brief description of our proposal} a new distributed
data-management scheme which is based upon Random Walk Based Membership Service
\cite{IEEEhowto:zbar} to facilitate Data Dissemination in Mobile Sink based Wireless Sensor
Networks.








\subsection{Moving Schemes for Mobile Sink in WSN \cite{IEEEhowto:yanzhong}:}

Basically in this paper, the author proposes a proactive algorithm to alleviate the Hotspots
problem by introducing a Mobile Sink to balance energy consumption among sensor nodes.

General Idea of Research Paper: Sink and sensor nodes know their own geographic location by either
using GPS or self-configuring localization techniques. WSN carries a neighbor discovery process.
Through neighbor discovery, sensor nodes can obtain the location information of their one-hop
neighbors. After neighbor discovery, sensor network starts gathering sensed data periodically. In
each data-gathering, the sensor nodes will send their data to the sink through multi-hop
communication path.

\subsubsection{Data Gathering Period}
It consists of three phases:\\

{\bf Phase 1:} Sink broadcast its position and all sensors received it and when sink changes its
position then sink calculates the distance from its current position to the old position where it informs
the whole network for the last time and sets its TTL field.\\

{\bf Phase 2:}
    Nodes select its neighbors as the next-hop to forward data packet. If the sink is in communication range,
    the sensor node will take the sink as its next hop. Each data packet carries the energy and position information
    of the node that have the highest residual energy and lowest residual energy among the nodes on its delivery path.\\

{\bf Phase 3:}
    The sink determines the direction and distance based on analyzing the energy distribution information carried by
    the data packets it received and then it moves to the new position before the next period begins.\\

\subsubsection{Moving Schemes}

{\bf 1. One-Step Moving Scheme}\\
This scheme is more suitable for networks that have fast moving sinks and that have long data-gathering intervals.\\

\noindent{\bf 2. Multi-Step Moving Scheme}\\
This scheme is suitable for networks where sink moves slowly and that have short data-gathering intervals.


\subsection{Coordinate Magnetic Routing for MSWSN \cite{IEEEhowto:shih}:}


In this paper the author introduces a new routing protocol for data dissemination in heterogeneous
WSNs.  They called their algorithm as Coordinate Magnetic Routing Algorithm. They also assumed that
their WSN is hierarchical sensor networks that have powerful nodes, which they called as CHs
(Cluster Heads), which have more resources compared to ordinary sensor nodes. These CHs are
responsible for relaying aggregated data to mobile sinks. These CHs are informed by mobile sinks
and these cluster head will relay the data to mobile sinks. These CHs (moles) can provide current
mobile sinks location and maintain routing paths according to sink movements.

\subsection{FLOW: An Efficient Forwarding Scheme to Mobile Sink in Wireless Sensor Networks \cite{IEEEhowto:rahul}:}


Forwarding using Likelihood-based Weights (FLOW) makes use of underlying pattern in the sinks'
movement to discover good delivery paths. They assumed that Moles are the nodes that lie in the
vicinity of the path that the sink takes. It is assumed that a Mole can somehow detect the presence
of the sink near itself. Every mole characterizes the sinks presence in its vicinity as a
probability distribution. These are then used by each node to calculate its likelihood of being on
a good path to the sink in a distributed fashion using only local information obtained from the
neighboring nodes. Forwarding decisions at each node are made using these values to send data to
the moles. They assumed a grid-based structure. The mobile sink moves periodically along the
trajectory.
\subsection{Data Dissemination to Mobile Sinks in Wireless Sensor Networks : An Information Theoretic Approach (WEDAS) \cite{IEEEhowto:habib}:}


Basically in this paper the authors proposed an energy aware protocol, called Weighted Entropy Data
Dissemination (WEDAS) for disseminating data to the mobile sink in WSNs, using an information
theoretic approach.

They also proposed that the selection of data disseminator should not only depend upon the
remaining energy  but also the distance. In this paper the authors tried to give a solution to
select static sensor nodes that will act as a data disseminator between sources and the mobile sink
based upon the position of the mobile sink and the remaining energy uncertainty of static sensors.
The WEDAS protocol favors sensor nodes whose weighted entropy with respect to their location and
remaining energy is the minimum to participate in building dissemination paths between sources and
the mobile sink.

They also prove that the total energy consumption in disseminating the monitored data to the sink
reaches its minimum when the data disseminator lies on the direct path between a source and a sink.








%
%

\section{Reactive approaches}
\label{sec:reactive}

Reactive Data Management and Data Dissemination Schemes does not requires any previous network
organization in terms of data management and data dissemination. The data is collected by the sink
during the period of visits.

\begin{figure}[H]
\centering
\setlength\fboxsep{0pt}
\setlength\fboxrule{0.5pt}
\includegraphics[height=60mm,width=120mm]{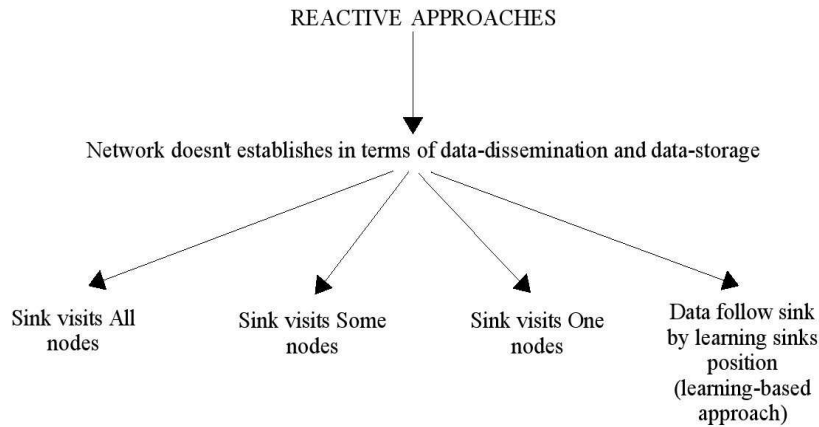}
\caption{Reactive schemes indicating the need of Mobile Sinks Visit} \label{fig_sim}
\end{figure}

In these schemes, several mechanisms can be used for defining the mobile sink's trajectory. These
mechanims depends upon the applied data dissemination approach but here the important issue that
need to be considered is that each approach should guarantee that the defined sink trajectory
should allow us to collect data from the whole network. For instance, a mobile sink may visit all
the nodes, visit some nodes, visit one node, or data follow sink by
learning sinks position \emph{(learning-based)} approach as depicted in Fig. \ref{fig_sim}.\\


For example in \cite{IEEEhowto:jun}, mobile sink has to visit all nodes by traversing the whole network and no distributed data management is performed.


\cite{IEEEhowto:guydong} is a good example, which shows that mobile sink has to visit some sensor nodes.


Another example is that if we let the mobile sink to visit only one sensor node in the network then it
is obligatory  that each node must collect data of all other remanding nodes. 


And finally when data follow sink by learning sinks position (learning-based approach) then it is
not obligatory that each node must collect data of all other remaining nodes. 



In reactive approaches when sensor nodes wants to send sensed data as a response to sinks query,
these sensor nodes either obtains the sinks location from some specific nodes who already knows the
sinks location or by using some techniques learns the position of the sink. We call this kind of
approaches, learning-based approach.

In reactive approaches mobile sinks sends some packets (query) and in reply the sensor nodes
establishes  the most feasible routes in terms of different criteria like less energy consumption,
high throughput etc., to disseminate sensed data towards mobile sink.

Another feature of reactive approaches is that the mobile sink selects some specific nodes in the network
to query data and provide its location to these specific nodes. In this way the mobile sink facilitates remaining
nodes that are in the vicinity of these selected nodes to send their data to these selected nodes and disseminate
data to the mobile sink. \\


At the following sections, we describe some reactive schemes. As previous described, our goal
here is not to provide a exhaustive description of related works in the literature, but to discuss
their general idea.

\subsection{Locators of Mobile Sink for WSN \cite{IEEEhowto:guydong}:}


Due to the high mobility of sink, geographic forwarding cannot be executed without a moving sinks'
location updates to the source or some forwarding nodes. Although one can think about periodical
flooding of sinks location to whole sensors but it is not efficient and scalable. To avoid this
problem, the author proposes a data dissemination model using geographic routing with locators to
support mobile sinks geographic routing. Locators are location server sensors that track sinks
current position and reply sinks' location query from sensors.

\subsection{Interest Dissemination with Directional Antennas for WSNs with Mobile Sink \cite{IEEEhowto:yihong}:}

In this paper the author proposed a directional-antenna-assisted reactive routing protocol to
resolve the problem of high packet loss rate and poor energy efficiency of traditional reactive WSN
routing algorithms caused due to frequent topology changes due to highly mobile sink nodes. IDDA
exploits the directional antenna to prearrange interest dissemination along the direction of
motion. In IDDA, with the prior knowledge of its velocity, the sink node uses a directional antenna
to broadcast interest packets along its direction of motion, and this rearranges an interest
dissemination in advance. When the mobile sink keeps moving along its orientation, IDDA collect
data back from sensor nodes in the vicinity reactively. If the prearranged interest dissemination
is carefully adjusted to a proper scale, the returning data will meet the mobile sink when it
arrive at the data aggregation point, increasing the packet delivery ratio and reducing the power
consumption. To improve IDDA's performance, they designed a cross-layer (PHY + NET) technique in
interest dissemination, which reduces energy consumption.

Furthermore, a power-aware dissemination algorithm is exploited and incorporated in IDDA to further reduce
energy consumption.
They compared IDDA and power-aware IDDA with Directed Diffusion in terms of energy dissipation per data report,
packet delivery ratio, and target detection ratio.\\

\subsection{Data MULEs: Modeling and analysis of a three-tier architecture for sparse sensor networks \cite{IEEEhowto:rc}:}

Basically in this paper the author neither proposes any sink trajectory nor improved any data
dissemination strategy. Instead in this paper the author focuses on a single analytical model for
understanding performance as system parameters are scaled. The performance metrics observed are the
data success rate (the fraction of generated data that reaches the access point) and the required
buffer capacities on the sensors and the MULEs.

In this paper the author wants to achieve cost-effective connectivity is sparse sensor networks while reducing the
power requirements at sensors.
The primary advantage of their approach is the potential of large power savings that can occur at the sensor because
communication now takes place over a short range.
The primary disadvantage of this approach, however, is increased latency because sensors have to wait for a MULE to
approach before the transfer can occur.
They does not address the issue of energy consumed during radio listening. This can be potentially high because a
sensor has to continuously listen to identify when a MULE passes by.\\

\subsection{MobiRoute : Routing Towards a Mobile Sink for Improving Lifetime in Sensor Networks\cite{IEEEhowto:jun}:}

Basically MobiRoute extends MintRoute \cite{IEEEhowto:aw} by adding functions that performs the following functions:

1.Notify a node when its links with the sink gets broken due to mobility.

2.Inform the whole network of the topological changes incurred by mobility.

3.Minimize the packet loss during the sink moving period.\\

\subsection{Efficient Data Propragation Strategies in WSNs using a Mobile Sink \cite{IEEEhowto:ioannis}:}

Basically in this paper the author propose the basic idea of having a sink moving in the network area and
collecting data from sensors.
They propose four characteristic mobility patterns for the sink that they categorized into two sub-categories
and they combine them with different data collection strategies.\\

\begin{enumerate}
\item Randomized
\begin{itemize}
\item Simple random walk.

\item Biased random walk.

\item Walks on spanning tree subgraphs.\\
\end{itemize}

\item Predictable

\begin{itemize}
\item Moving on a straight line or cycle.\\
\end{itemize}
\end{enumerate}

To get data from sensors, the sink movement is combined with three data collection strategies:

\begin{enumerate}
\item Passive data collection strategy.

\item Multi-hop data collection strategy.

\item Limited multi hop data collection strategy.\\
\end{enumerate}

There are many different approaches when considering the mobility pattern that the mobile sink should follow.\\

\begin{enumerate}
\item Random Mobility:

The movement of the sink is done in a random manner regarding the position and the speed of movement. The main characteristic of this pattern is simplicity and unpredictability of the future position of the sink.

\item Predictable Mobility:

The movement of the mobile element follow a certain pattern that can be computed. Such movements may be periodic movements along a predefined trajectory.

\item Controlled Mobility:

The mobile element can vary its movement in a deterministic way in order to achieve better results.\\
\end{enumerate}

They propose the following four protocols.

\subsubsection{Protocol \# 1: Random Walk and Passive Data Collection}

The simplest of all possible mobility patterns is the random walk, where the mobile sink can move
 chaotically towards all direction at varying speeds.
Data is collected in passive manner.
Sensors cache all recorded data,
periodically a beacon message is transmitted from the sink.
Each sensor node that receives a beacon attempts to acquire the medium and transmit the cached data to the sink.
Transmitted data is then removed from sensor’s cache to free memory for new readings.\\

\subsubsection{Protocol \# 2: Partial Random Walk with Limited Multi hop Data Propagation}

Another form of Random Walk is performed by using a set of predefined areas and random transitions between the areas
according to their connectivity.

\subsubsection{Protocol \# 3: Biased Random Walk with Passive Data Collection}

The idea of using a logical graph can be extended in a way that certain areas of the network are
favored (i.e. more frequently visited) by the sink in order to improve the data collection process
or to overcome problems that arise form the network topology.
The selection of the next area to visit is done in a biased random manner depending on these two variables:

\begin{enumerate}
\item Frequency Biased: In this, less frequently visited areas are more likely to be visited when the sink
 is located at a nearby area.

\item Density Biased: In this, areas with many sensor devices are more likely to be visited when the sink is
located at a nearby area.\\
\end{enumerate}

\subsubsection{Protocol \# 4: Deterministic Walk with Multi hop Data Propagation}

Here they use a single form of controlled mobility where the mobile entity moves on a predefined
trajectory. They examine the cases where the trajectory is a line $M_{line}$ or a circle
$M_{circle}$ that is fully contained in the network area. The trajectory is characterized by its
length l. In particular, the linear trajectory consists of a horizontal or vertical line segment
passing through the center of the network. The sink moves from one edge of the line to other and
returns along the same path. For the case of a circular trajectory, the circle is centered at the
center of the network. Initially the sink is positioned on the circumference of the circle and
continues along this path.

\pagebreak
\section{Our Contribution and Proposal}
\label{sec:ourproposal}

In this report, firstly, we have presented state-of-the-art survey on Data Management and Data
Dissemination techniques with Mobile Sink. Moreover, we classified these techniques into two ample
sub-categories. Under this classification, we identify, review, compare, and highlight these
techniques and their pros and cons. We have done a SWOT (Strength, Weaknesses, Opportunities,
Threats) analysis of each scheme. We also discussed where each scheme is appropriate.

Secondly, we are now presenting a new scheme which is based upon Random Walk Based Membership
Service to allow Data-Dissemination in Mobile Sink based Wireless Sensor Networks. A membership
service provides each node with a view regarding who are the other nodes in the network.


\subsection{Introduction to our proposal}

Mobile Sink based Wireless Sensor Networks introduces may challenges like distributed storage
capability, specially how to safely store collected sensed data so that it can be retrieved later
and besides this, energy optimization of
 Wireless Sensor Network Nodes is also another important requirement.

To address these challenges, we present here our proposal for efficient data dissemination and data storage in
Mobile Sink based Wireless Sensor Network. Our main goals here are:

\begin{enumerate}

\item to optimize energy consumption by aggregating collected data in some selected storage nodes, and

\item To improve data availability by replicating aggregated data in selected storage nodes.\\

\end{enumerate}

\subsubsection{Problem statement}
The problem addressed in this report is threefold: (i) to reduce the aggregation
cost by limiting the distance d from any mote node to its storage motes; (ii)
to determine the degree of replication k required (iii) to limit the energy con-
sumption of involved storage nodes by limiting the number k of storage node
per mote.

\begin{enumerate}                                      
\item \textbf{Aggregation cost.}
Aggregation is a fundamental issue in wireless sensor
networks, however, it can become costly if the storage mote is far from the
sensed event location. Consider for any mote i, the aggregation cost as the
distance d between node i and the storage mote where it has to transmit the
collected information. Using this definition, limiting the aggregation cost over
the system aims at limiting the largest distance d from any mote to its storage
motes.
\item \textbf{Replication degree.}
 Due its resource limitations, wireless sensor networks are inherently dynamic. The only way to cope with dynamism requires
periodic replacement of motes. Nevertheless, between two mote nodes replacement, motes must cope with fault tolerance to provide data persistence despite
failures. In order to determine the cost of replication one must determine the
number k of storage motes that will replicate the data.
\item \textbf{Energy saving.}
 For accessing data the mobile sink has to contact at
least one storage mote of each mote. Communication between the mobile sink
and the storage motes implies that only these motes can be active. This translates into a signiﬁcant energy waste if all network mote nodes are active. In this
way, the number of storage motes has to be determined not only to guarantee
data reliability, but also to limit the number of active motes in the network,
limiting thus, the energy consumption.\\
\end{enumerate}
    Actually, the adopted solution to the data dissemination over the WSN—a
consequence of the combination of aggregation costs, replication degrees, and
level of energy savings, directly impacts the trajectory that the mobile sink
should (or may) take through the deployment region. In some cases, the management of the trajectory of the mobile sink should be integrated into the adopted
solution as it determines the sensor nodes to be visited in order to gather a representative view of the monitored field. Whereas in other cases some distributed
solutions may allow the mobile sink to use any trajectory as long as it visits a
certain number of sensor nodes to gather such a representative view.
\subsubsection{Problem complexity}
Determining a good tradeoff between the previously described parameters,\emph{ i.e.
d and k} is a complex task. Some naive solutions could be:
\begin{enumerate}
\item \textbf{All nodes are storage motes, being no aggregation performed:}
considering \emph{d}, and \emph{k}, this solution consists to take \emph{d} = 0, and \emph{k} = 0. This configuration requires the visit by the mobile sink, of all nodes (see Fig.~\ref{fig:config-a}).
Nevertheless, even if it can be considered that mobile sinks has the required
resources to visit all mote nodes in the monitored area, this configuration \emph{requires a longer time for data retrieval, results in a higher consumption of motes                                        
resources, imposes limits for data collection by motes, and finally, has no fault
tolerance}. These are due the fact that mote nodes have to:
\begin{itemize}
     \item become awake during all the retrieval period;
     \item keep collected data in the memory until their sink retrieval, which can
       restrict new data collection in the case of memory resource limitation,
       and
     \item nodes represent points of failures, since no replication is performed.\\
\end{itemize}

\begin{figure}[htbp]
  \begin{center}
    \subfigure[]
    {
      \label{fig:config-a}
      \epsfxsize=4cm
      \leavevmode\epsfbox{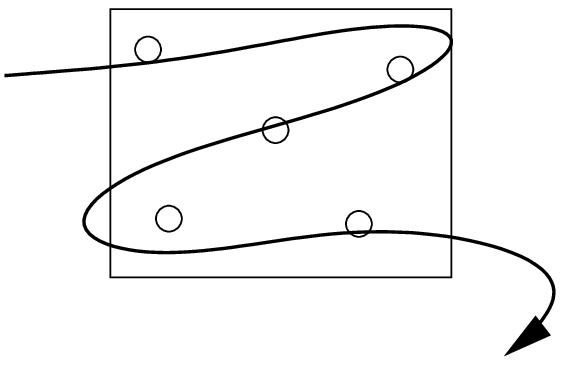}
    }\hspace{1cm}
  \subfigure[]
    {
      \label{fig:config-b}
      \epsfxsize= 2.5cm
      \leavevmode\epsfbox{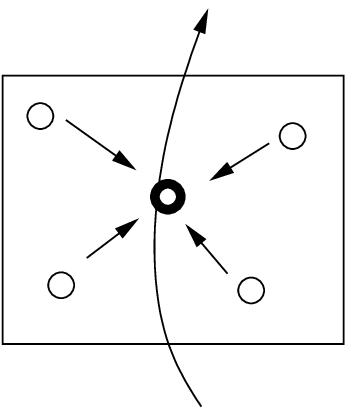}
    }\hspace{1cm}
    \subfigure[]
    {
      \label{fig:config-c}
      \epsfxsize= 2.5cm
      \leavevmode\epsfbox{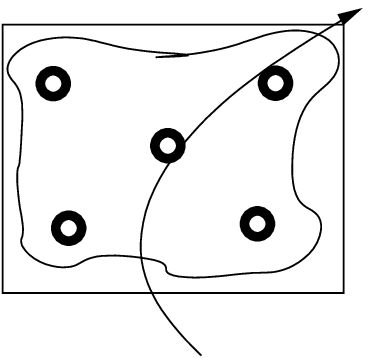}
    }
    \caption{Data retrieval by a mobile sink. (a) No aggregation is performed and mobile sink has to visit all mote
    nodes to retrieval their data: $d=0$, $k=0$. (b) One mote aggregate all collected data and mobile
    sink has only to visit that node: $d=${\it (network-diameter/2)} and $k=1$. (c) Each mote is a storage mote for
    all the $n-1$ remaining motes in the network:
    $d=${\it network-diameter}, and $k=${\it number of motes}.}
    \label{fig:config}
  \end{center}
\end{figure}


\item \textbf{One centered node plays the role of storage mote and aggregates data from all the other nodes:}
considering \emph{d} and \emph{k}, the solution would be
\emph{d = (network-diameter/2)} and \emph{k = 1}. In this case, a single storage mote would
represent a centered node in the topology aggregating data from all other nodes.
This requires the visit of an unique node by the mobile sink (see Fig.~\ref{fig:config-b}).
Nevertheless, even if it reduces the trajectory length of mobile sinks and the
data retrieval time, this configuration \emph{decreases the amount of information that
can be retrieved, imposes a higher communication overhead, strongly reduces the
lifetime of the selected storage mote and its neighbors, and ﬁnally, has no fault
tolerance}. These are due the fact that:
\begin{itemize}
     \item the storage mote has limited memory, which can restrict the storage of
       new received data;
     \item all collected data has to be sent to the storage mote, generating high
       communication load to the network;
     \item being the storage mote a hot spot, its neighborhood nodes will also have
       their resources affected;
     \item the storage mote represents a single point of failure.
\end{itemize}

\item \textbf{Replication is performed on all nodes in the network: }
considering
\emph{d} and \emph{k}, the solution would be \emph{d =size-network}, and \emph{k =number of motes}.
In this case, each mote is storage mote for the n-1 remaining motes in the
network, being the replication cost unaﬀordable. Thus, data collected by any
mote is aggregated by all storage motes. This requires the visit of at least one
mote in the network by the mobile sink (see Fig.~\ref{fig:config-c}) to retrieval any collected
data. Nevertheless, even if it assures a high fault tolerance and reduces the
trajectory length of mobile sinks and the data retrieval time, this configuration
\emph{decreases the amount of information that can be retrieved and results in a higher
consumption of motes resources}. These are due the fact that:
\begin{itemize}
     \item become awake during all the retrieval period, and                                        
\item keep collected data in the memory until their sink retrieval, which can
  restrict new data collection in the case of memory resource limitation;
\item imposes a high replication cost.
\end{itemize}

\end{enumerate}

In our proposed scheme, the main problem we tackle in the context of Mobile Sink
Wireless Sensor Networks (MSWSNs) is how to make the monitored data available to the mobile sink in
a robust, adaptive, and safe way. In short, we investigate an efficient data dissemination
approach (in terms of overhead and representativeness) to allow a mobile sink to gather a
representative view of the monitored region covered by $n$ sensor nodes by visiting any $\sqrt{n}$
nodes. We thus propose our scheme using Random Walks (RWs) with
uniformly distributed views to disseminate data through the WSN with a controlled overhead.

The proposed approach does not require a priori knowledge of all network nodes, does not use
multi-hop routing or any sink's track mechanism. Only one mobile sink is required, being this one mobile sink
free to follow any trajectory. Finally, our approach improve data availability by replicating
aggregated data in selected storage nodes in the network.

The general idea is to set each storage mote with a \emph{view} defined as a set of
node descriptors:\\

 $<$NodeIdentifier, DataValue, LastTime$>$ \\

Where $\emph{NodeIdentifier}$ is the ID of the source mote, $\emph{DataValue}$ refers to the
monitored data in this particular sensor at given time, and $\emph{LastTime}$ is the last time the
storage mote has heard from this source. The source then advertises itself every \emph{U} time
units by starting a reverse sampling process. Thus, a RW is started at the source by randomly
selecting the next-neighbor to send the message, until the distance d to be reached. That \emph{U}
time needs to be set according to the mobile sinks' visits, which might also be defined by the
application and/or the sensors memory.

Consider that the typical intersection between views of neighboring nodes results from ideal uniformly chosen views
and thus there is no special correlation
between the views of neighboring nodes, as discussed in the RaWMS paper.
This uniform view distribution achieves an average view size of $\sqrt{n}$ with an expected intersection of
\begin{equation}
{\sqrt{n}}\frac{\sqrt{n}}{n} = 1,
\end{equation}
for all network sizes, when adopting a sufficiently large walk length (n or n/2). Based on the uniformly distributed and
little correlated information offered by RaWMS, one may expect neighboring
nodes to have little information intersection between them, i.e. a mobile sink
might obtain a representative view of the information concerning the deployment region by only visiting a
relatively small number of sensor nodes in the
network. Thus, considering the previously defined parameters (i.e., the distance
d of each random walk and the k storage nodes), d = n or d = n/2 , and k = $\sqrt{n}$.

Indeed, with the uniform distribution of information and the small intersection between neighboring
nodes, each visited sensor node adds a large amount of uncorrelated information to the mobile
sinks' view of the whole monitored field. More formally, as the mobile sink visit a sensor node, it
gathers data from $( \sqrt{n} - 1)$ nodes. As the mobile sink keeps visiting other nodes, at the
$i^{th}$ sensor node it has collect approximately data from $i( \sqrt{n} - 1)$ nodes. Therefore,
based on the uniformness provided by RaWMS and the little intersection of the stored information at
sensors, one may expect that the mobile sink would be able to get a representative view of the
monitored field (i.e. information about a number of sensor nodes close to n) by only visiting any i
$\approx$ $\sqrt{n}$ different nodes following a random trajectory through the deployment region.

This strategy has the clear advantage of leaving the mobile sink free to follow any trajectory
through the deployment region, thus decoupling the data dissemination management from the
management of the mobile sinks' trajectory. In other words, the mobile sink should only be
concerned in visiting a certain minimal number of nodes to achieve a representative view of the
monitored field, no matter which nodes. The all procedure will be then re-started after \emph{U}
time after mobile sinks have collected all data.

In our proposed scheme, we are assuming that each node in the network has a membership view
regarding some other nodes in the network. We are not going to assign complete membership view
because again it will increase the storage capacity of nodes which is again a constraint. Instead,
we are suggesting an optimized random view of the network membership based upon RaWMS (Random Walk
based Membership Service) sampling technique \cite{IEEEhowto:zbar}. This membership service
provides a random chosen partial membership view for nodes in the network, by guaranteing  the
uniformness of the location of nodes appearing in the views.

The general idea is that initially all wireless sensor nodes follows an Active/Sleep Regime where
they sleep a fraction $ 0 \leq \delta \leq 1$ of each unit of time \emph{U}. In simple words each
sensor node will active for $t_a$ seconds before it sleeps for $t_s$ seconds, in order to save
energy. By active state, we means that a sensor node turns on its radio, while sensor node in sleep state means that it will just only turn off its radio. In this case,

\begin{equation}
U = t_a + t_s   \\
\end{equation}

\noindent and,

\begin{equation}
\delta = \frac{t_s}{t_a + t_s}
\end{equation}

We also assumes that there is no synchronization among different sensor nodes with respect to their regime of
activity as each sensor starts following its Active/Sleep regime after the expiration of a timer set to be
uniformly distributed over an interval equivalent to \emph{U}.

Therefore, considering a MSWSN composed of \emph{n} nodes, $[(1-\delta)n]$ sensor nodes are expected to be active
on average at any given time equal to U.

Then during their activity regime each wireless sensor node generate Random Walks at each \emph{U} intervals of
time. Thus, a Random Walk is started at the source by randomly selecting the neighbor to send the message until
the distance \emph{d} to be reached. Then mobile sink will visit $\sqrt{n}$ nodes and in this way it will collect
stored sensed information of n sensor nodes.

\subsection{Advantages of Our Proposal}

The Advantages of our proposal are:
\begin{itemize}

\item    No need of multi-hop routing.
\item Mobile Sink is free to follow any trajectory. In simple words, the Mobile Sink should
    only be concerned in visiting a certain minimal number of nodes to achieve a
    representative view of the monitored field, no matter which nodes.
\item Improved data availability and by consequence, fault tolerance,
by adopting the feature of data replication (storage motes).

\item Active and sleep states of sensor nodes, resulting in energy saving thus increased in
    network lifetime.
\item Apart from sensor nodes, storage motes can also be in active and sleep state, thus saving energy and
increases network lifetime.
\item  Proactive in nature. First, network is configured keeping data management and data
    replication in consideration and then Mobile Sink can collect data by visiting only a
    limited number of motes, i.e. only $\sqrt{n}$ nodes.
\item  Eliminate the hotspots problem.
\item  No need, but possible, of Multiple Mobile Sinks.
\item  By providing a membership view by using RaWMS (Random Walk Based
Membership Service), our proposal solve the problems generated when: {\it (a)} all nodes are
storage motes, being no aggregation performed; {\it (b)} one certain node plays the role of storage
mote and aggregates data from all the other nodes;{\it (c)} replication is performed on all nodes
in the network.

\end{itemize}
\subsection{Comparision between Our Proposal and State-of-the-art Data-Dissemination and Data-Management Techniques}

In this section we are describing a detail comparison between our proposal and State-of-the-art
Data Dissemination and Data Management Techniques. Apart from it, we are also discussing
differences of our proposal over different schemes:

\begin{itemize}
\item Moving Schemes for Mobile Sinks in Wireless Sensor Networks \cite{IEEEhowto:yanzhong}:

This scheme does not support replication of  data. Our proposal has an advantage which is that
neither sink is responsible for broadcasting its position nor sensor nodes are responsible for
sending their position information. 

\item Locators of Mobile Sinks for Wireless Sensor Networks \cite{IEEEhowto:guydong}:


Basically in this paper the authors are assuming that MSWSNs require an additional mechanism of
geographic routing and sinks location should be propagated continuously resulting in the drain up
of the sensors battery power and increase wireless channel contention. But as far as our proposal
is concerned, we are not supposed to utilize any sort of geographic routing. Evidently the feature
of not having this capability will increase the network lifetime. The second thing that is
different is that our proposal has a replication mechanism which is an advantage in terms of data
while this proposal proposed replication of locators (locator failure) instead of data replication.
The data generated by the sink to update certain locators is more than compared to our solution.
Again sensors also has to contact the locators to find the sink position and thus more traffic is
generated and thus, more energy consumption. This locator protocol also needs another geographic
based routing protocols like GPSR or greedy forwarding. In this paper, sink has to visit a certain
number of locators and thus increase the mobility of sink, while in our proposal sink has to visit
just a limited number of nodes. We also proposed active and sleep state of sensors while they did
not proposed it. Their proposal do not eliminate the Hotspots problem. Locators are present in a
grid and there are certain number of sensor nodes that will contact them, resulting in Hotspots of
locators.

\item Coordinate Magnetic Routing for Mobile Sinks Wireless Sensor Networks \cite{IEEEhowto:shih}:

The main advantage between our proposed proposal and their proposal is that their proposal contains
a heterogeneous WSN. By this, we mean that they have specialized Cluster Head (CH) nodes that have
more power then ordinary sensor nodes.
The main disadvantage of their protocol is that in their proposed protocol all the nodes have to
calculate the distance resulting in more energy consumption.
In this paper, they tried to improve the data dissemination method and they also tried to improve
the sink's trajectory, but, delay factor in collection of data is involved. Sink have to
broadcast hello messages to the neighboring Cluster Heads to notify its location thus resulting in
energy consumption and reducing the network lifetime.


\item Data Dissemination to Mobile Sinks in Wireless Sensor Networks: An Information Theoretic Approach \cite{IEEEhowto:habib}:

One disadvantage of their proposal is that the sensors should be aware of their fixed position.
Sensors advertise their position to their neighbors at the start of the monitoring task, resulting
in more energy consumption, which is another disadvantage. They advertise their remaining energy to
their neighbors by piggybacking it on the monitored data sent to the sink, thus also resulting in
energy consumption of sensors. Since sink has to advertise its current and future positions to its
neighbors, thus this is also another disadvantage and which is not present in our solution. Since
in WEDAS protocol, every sensor node has to keep track of its energy then this is also a
disadvantage.

\item FLOW: An Efficient Forwarding Scheme to Mobile Sink in Wireless Sensor Networks \cite{IEEEhowto:rahul}:

Their proposal is based upon multi-hop routing and thus resulting in more energy consumption.
One disadvantage is that there must be a particular sink movement pattern and that pattern must be
sense by the nodes (sink) and statistically characterize it as probability distribution function
(PDF). Mobile sink will not query any data request, instead it is the sensor node, who when sensed
data, calculate the route and then send data to the mole, which is in the vicinity of mobile sink.
Mobile sink will then receive that data.

\item Sensor Network Calculus with Multiple Sinks \cite{IEEEhowto:jens}:

Basically in this paper the authors neither propose a sink trajectory nor any sort of optimization
of data-dissemination. Instead, the author showed that how sensor network calculus may be able to
shed some light upon how the number of sinks affects the worst case message transfer delay in
typical WSN. So, our proposed idea and their idea are not similar.

\item MobiRoute : Routing Towards a Mobile Sink for Improving Lifetime in Sensor Networks \cite{IEEEhowto:jun}:

Their algorithm adaptively changes the sojourn time of the sink at each anchor point according to the power
consumption profile of the network.
One disadvantage of their proposal is that they considered a scenario where nodes of a WSN periodically sample
data and transfer these data through multi-hop routes towards the sink.
Another disadvantage of their proposal is that sink spends energy to send s-beacons, nodes also spend energy to
receive these beacons, resulting in more energy consumption and less network lifetime.
One disadvantage is that nodes can buffer data packets resulting in energy consumption of the sensor nodes.
They used two types of packets in their proposed protocol:
Control packets.
Data packets.
Both the above types of packets resulting in energy consumption.
Since during each sampling period, the sink collects the power consumption records for all nodes and thus
resulting in less network lifetime.
The main disadvantage is that power is consumed while collecting power consumption data.

\item Data MULEs: Modeling and analysis of a three-tier architecture for sparse sensor networks \cite{IEEEhowto:rc}:

In this paper the author focuses on a simple analytical model for understanding performance as
system parameters are scaled. The performance metrics observed are the data success rate and
required buffer capacities on the sensors and the MULEs. 
Another characteristic is that, in their approach, however and increased latency because sensors have to wait
for a MULE to approach before the transfer can occur. They just considered Random Walk Mobility
pattern which is a restriction in terms of mobility. They did not discussed the active and sleep
states of sensor nodes.

\item Efficient Data Propagation Strategies in Wireless Sensor Networks using a Mobile Sink \cite{IEEEhowto:ioannis}:

One disadvantage of their proposal is that the sensors are not in sleeping state and thus resulting in more energy
consumption.
Since the sink must know its position, so it is also a disadvantage.\\

While considering \emph{Random walk and passive data collection} protocol: No network knowledge at
all is assumed, which is an advantage. This is a reactive protocol in which sink periodically send
a beacon message attempts to acquire the medium and transmit the cached data  to the sink. This
method may lead to many collisions, when the sink visits a dense area, which is a disadvantage.
Another disadvantage is that, since only a single transmission per sensed event is performed, it
minimizes energy consumption, but on the other hand, time efficiency may drop due to long intervals
between visits to the sensors. Since sensor cache all recorded data so there is more energy
consumption. Since transmitted data is then removed from sensors cache to free memory for new
recordings so there is no
fault tolerance in terms of data loss.\\

While considering \emph{Partial Random Walk with limited multi-hop data propagation} protocol: This
protocol is a reactive protocol because a graph is constructed during a graph formation phase that
is executed by the sink during the network initialization. This protocol depends upon multi-hop
routing, which is a disadvantage. This is again a reactive protocol. In this protocol, since sink
has to frequently create propagation trees, this generates an overhead and thus, consumes energy.
Since this protocol assumes and uses more knowledge of the network, it is also more expensive in
terms of communication and computational cost on the sensor devices.\\

While considering \emph{Biased Random Walk with passive data collection} protocol:
This protocol uses knowledge collected by the sink in order to speed up the coverage of new areas (when alpha
< beta), with an increase in computational overhead of the sink.\\

While considering \emph{Deterministic Walk with multi-hop data propagation} protocol: Since the
mobile sink covers only a small network area, it is necessary to collect data with a multi-hop data
propagation protocol, which is a disadvantage. The deployment of this protocol imposes a high cost
on the sensor devices that performs tree formation and multi hop propagation which is also a
disadvantage. In their simulation, they did not consider the possibility of nodes failures which is
another disadvantage.

\item Interest Dissemination with Directional Antennas for WSNs with Mobile Sinks \cite{IEEEhowto:yihong}:

Our proposal has many advantages over their proposal like:
No need of multi hop routing.
Concept of storage motes.
Sink is free to follow any trajectory etc.\\
One disadvantage of their proposal is that the mobile sink node has prior knowledge about its
trajectory. The advantage of their proposal is that it is not necessary for the mobile sink to
know locations of other sensor nodes. For transmission, it is necessary for mobile sink to use a
directional antenna and its receiving antenna should be omnidirectional.
One disadvantage is that in IDDA, with the prior knowledge of its velocity, the sink node uses a directional
antenna to broadcast interest packets along its direction of motion, and this rearranges an interest dissemination
in advance.

\end{itemize}

\pagebreak
\subsection{Simulation Results}
\subsubsection{Analyzing the behaviour of Active/Sleep Regime}

In this section, we are presenting our Simulation Results. For our simulation, we selected NS-2
Simulator version 2.32. Each simulation lasted for 500 seconds (simulation time) and each data point is
generated for an average of 15 simulation runs. Nodes uses IEEE 802.11 protocol for MAC layer and the propagation
channel used was Two Ray Ground. We used Uniform Random Numbers provided by NS-2. We then used these Random Numbers
for timeout time of our Wireless Sensor Nodes and that are with the range of 0 to 10. \\

We presents the algorithm for a sensor node in Appendix 1. A new RW is initiates after every \emph{U} time units and we used Hello Packets for neighbour discovery. HandleRW() method handles the RW. The publishView() method is reposnbile for managing timebased or sizebased Views. PickNextNode() method is responsible for picking the next node, while checking its status.

\subsubsection{Bonn Motion}
The Wireless Sensor Nodes were placed at Uniform Random Locations in a flat grid of 1000x1000 for analyzing
the behaviour of \emph {Active/Sleep Regime}. For this we used Java Based \emph{BonnMotion : A Mobility Scenario
and Analysis Tool}, prepared by University of Bonn, Germnay, to uniformly and
randomly place our Wireless Sensor Nodes in a flat grid.\\

\subsubsection{Graphs on Active/Sleep Regime}

In this section, we are presenting the graphs of Active/Sleep regime that we have obtained after
extensive simulations of NS-2.

\begin{figure}[H]
\centering
\vspace{-100pt}
\includegraphics{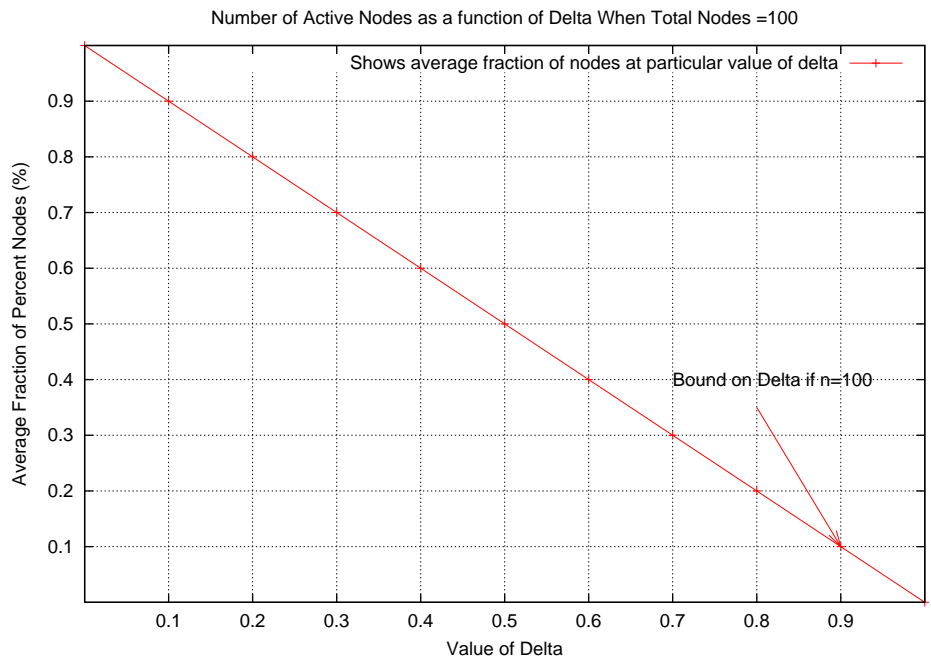}
\vspace{40pt}
\caption{Number of Active Nodes as a function of Delta }
\label{fig_sim10}
\end{figure}

Fig.~\ref{fig_sim10} represents the number of active nodes as a function of $\delta$ when there are
total 100 nodes. All these simulations were run on 100 nodes. We considered a unit time \emph{U =
10s} to extract the average number of active nodes as a function of an increasing $\delta$ (from 0
to 1). The objective here is to determine the upper bound on the $\delta$ fraction of nodes that
may be put to sleep, in order to guarantee at least $\sqrt{n}$ nodes awaken averagely in the network at any given time. \\

In simple words, we can say that Fig.~\ref{fig_sim10} tells us that if we want $\sqrt{n}$ nodes to
be in Active State on average at any given time equal to \emph{U}, then we should select
$\delta=0.9$. It means that if for 100 nodes, we put each node in Active State for 1s and Sleep
State for 9 sec. We then found there are at least $\sqrt{n} = \sqrt{100} = 10$ nodes, which are
Active in the network on average at any given time equal to \emph{U= 10s}.

\pagebreak


\begin{figure}[H]
\centering
\vspace{-120pt}
\includegraphics{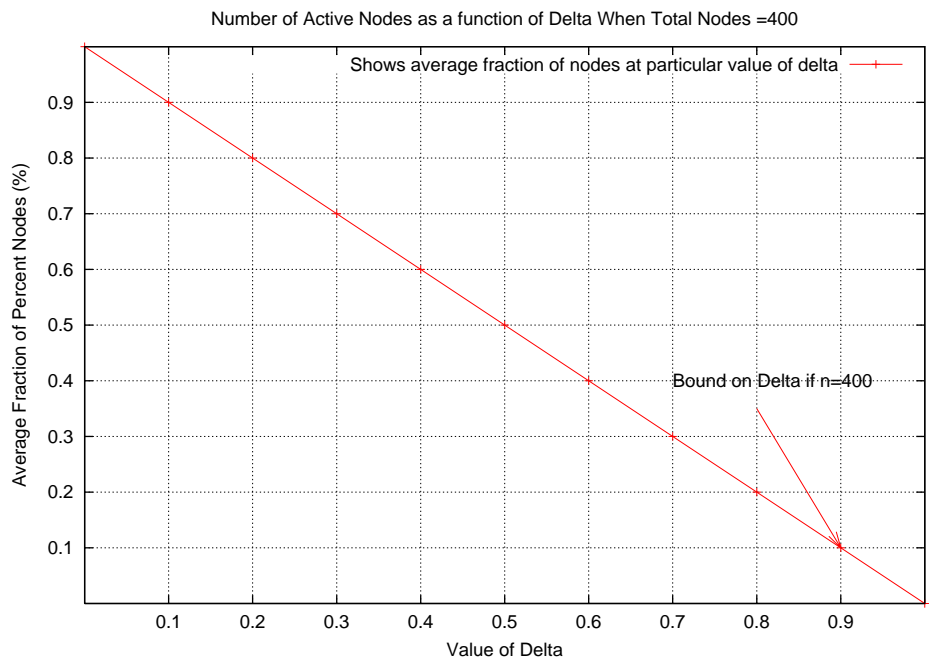}
\vspace{40pt}
\caption{Number of Active Nodes as a function of Delta }
\label{fig_sim17}
\end{figure}

Fig.~\ref{fig_sim17}. represents the number of active nodes as a function of $\delta$ when there
are total 400 nodes. Here, we found that for \emph{$\delta = 0.9$}, we found 40 nodes in the Active
State. Since we need at least $\sqrt{n} = \sqrt{400} = 20$ nodes in Active state in the network on
average at any given time equal to \emph{U= 10s}, so it is satisfying our needs.



\begin{figure}[H]
\centering
\includegraphics{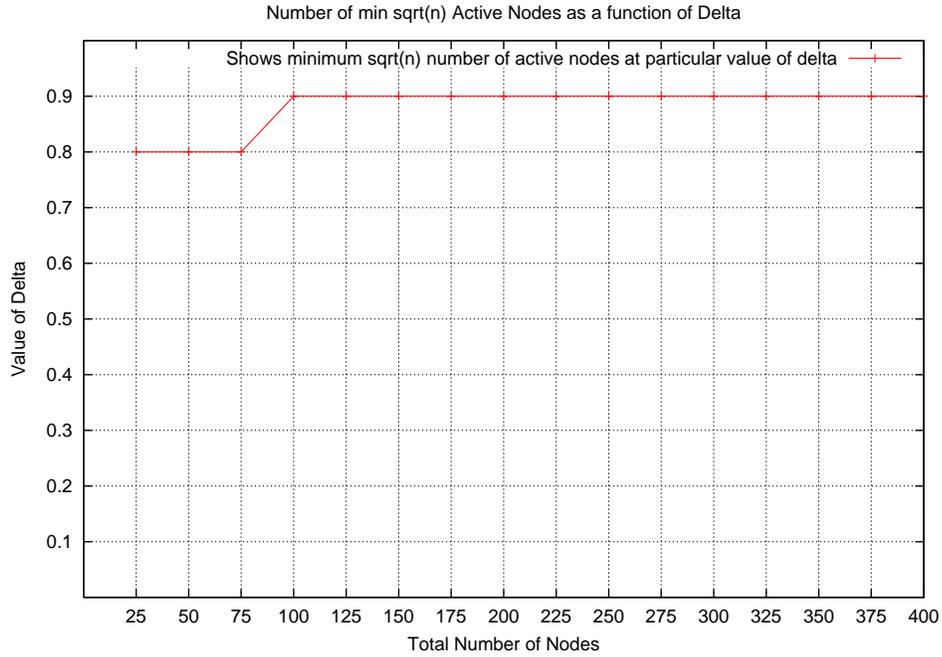}
\vspace{40pt}
\caption{Number of Minimum sqrt(n) Active Nodes at particular value of Delta }
\label{fig_sim13}
\end{figure}


Fig.~\ref{fig_sim13} shows the value of $delta$ required to get a minimum number of $\sqrt{n}$
active nodes for different network sizes. All these simulations were run on a network of size
varying from 25 to 400 nodes. Through this Fig.~\ref{fig_sim13}, we can deduce that if one wants
$\sqrt{n}$ nodes to be in Active State when there are total 300 nodes, then one should set the
value of $\delta = 0.9$ (considering \emph{U=10s} , $t_a=1s$  and $t_s=9s$).

\pagebreak


%
%
%
%

\begin{figure}[H]
\centering
\includegraphics{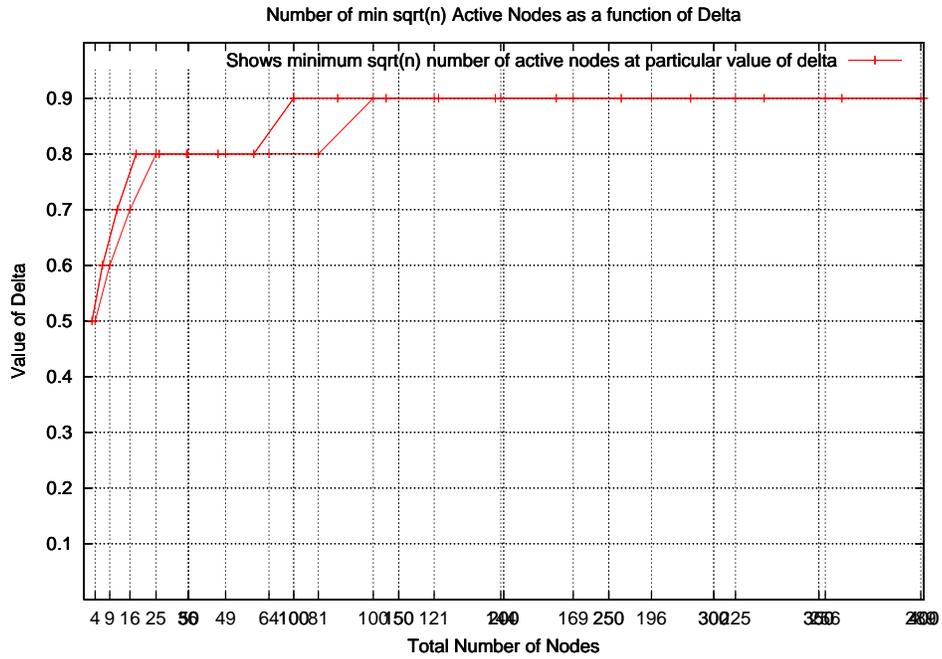}
\vspace{40pt}
\caption{Number of Minimum sqrt(n) Active Nodes at particular value of Delta }
\label{fig_sim16}
\end{figure}

Fig.~\ref{fig_sim16} also shows the value of $delta$ required to get a minimum number of
$\sqrt{n}$ active nodes for network sizes varying from 4 to 289 nodes.

\pagebreak
\subsubsection{Analyzing the behaviour of Random Walk}

In this section, we are analyzing the behaviour of data dissemination using RWs \emph{Random
Walks}. This will consist in making sink to visit $\sqrt{n}$ nodes and verifying the information of
how many nodes were really collected. Here, we are assuming that the interval of mobile sinks' visits should be at least
equivalent to U. New Random Walks may be then, generated at each U intervals of time, allowing new
collected values' and energy consumption's distribution among nodes in the network. Note that this
configuration is equivalent to a worst-case scenario where the sink stays in the region of
deployment for the minimal interval time to gather a representative view of the monitored field.

\subsubsection{Graphs on Random Walk}

\begin{figure}[H]
\centering
\includegraphics{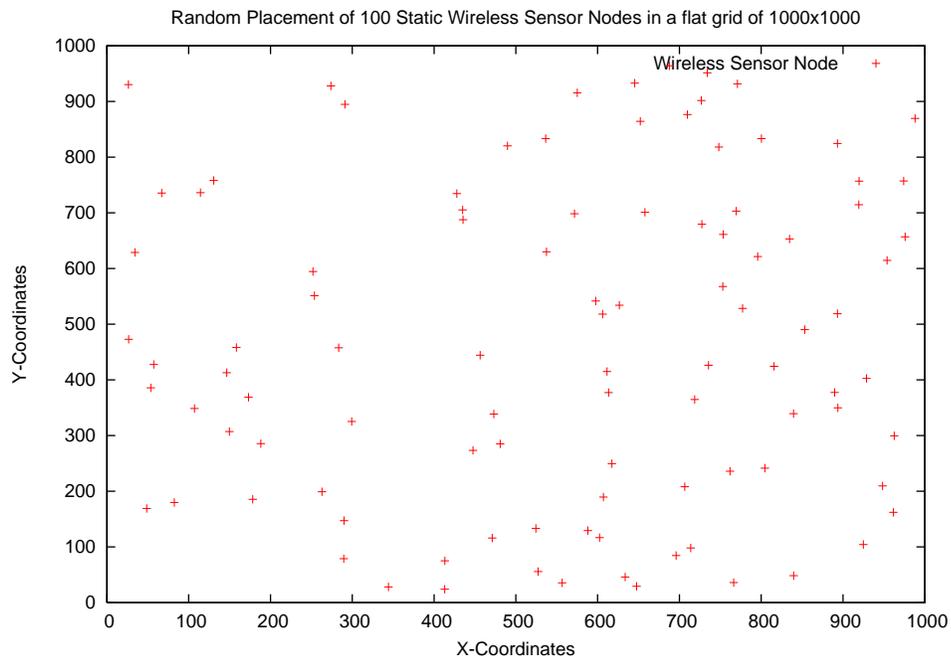}
\vspace{40pt}
\caption{Static Wireless Sensor Nodes placement on a flat grid of 1000x1000}
\label{coordinates}
\end{figure}

Fig.~\ref{coordinates} shows 100 wireless sensor nodes which were placed on uniform random locations on a flat gird of 1000x1000, which we obtained from \emph{Java Based Bonn Motion} and the network is fully connected.

\pagebreak

\begin{figure}[H]
\centering
\includegraphics{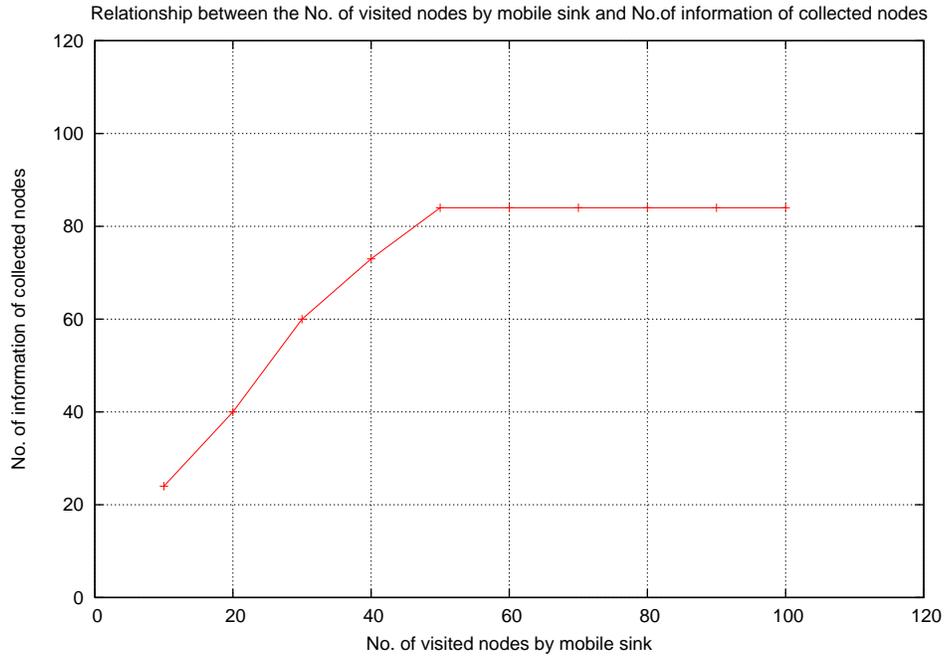}
\vspace{40pt}
\caption{Relationship between the No. of visited nodes by mobile sink and No. of information of collected nodes with normal parameters}
\label{rwnormal}
\end{figure}

Fig. ~\ref{rwnormal} shows the results, which we have obtained to analyze the behaviour of RW. For the simulation \footnote{While analyzing RW, due to lack of time i did not generate many experiments for each showed result.}, we assumed 100 static wireless sensor nodes. To analyze the behaviour of RW, we run the simulation for 1000s (of simulation time). Here we consider Random Walk length of \emph{n/2} and we are limiting the view size of each node, that is, we are adopting size based management of views of each node (\emph {$\sqrt{k}$}), and mobile sinks follow a random trajectory and randomly select any node and collects data.

In Fig. ~\ref{rwnormal}, we observed that there is a great impact of long timeouts on the data dissemination, and consequently, on the data collection. Here, we considered randomly generated time out states of nodes that are between 1 and 10 and we found that with this configuration, mobile sink is only able to collect very less representative view of the network.

With a high range of timeout at the initialization phase, nodes have a high probability to be active in distinct intervals
during the simulation duration. This decrease the chances of a node to find awake storage nodes in the network, decreaing the size of views.

\begin{figure}[H]
\centering
\includegraphics{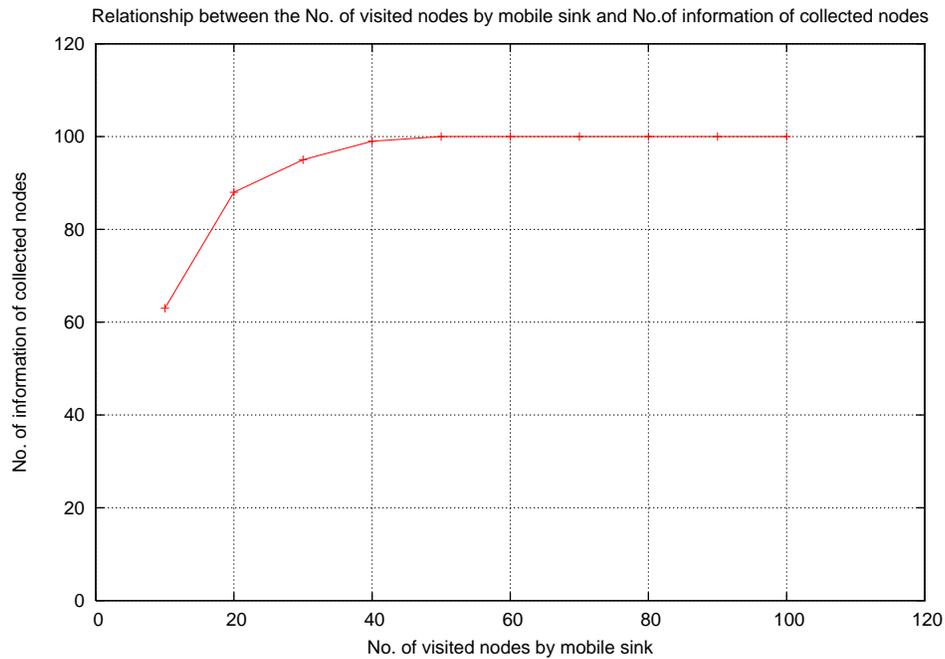}
\vspace{40pt}
\caption{Relationship between the No. of visited nodes by mobile sink and No. of information of collected nodes with very small time out}
\label{rwatimout}
\end{figure}

To analyze the behaviour of time out states of sensor nodes on the data dissemination, we took very small time out states of sensor nodes and found that a mobile sink is able to get the representative view of the whole network by just visiting very few nodes.

\begin{figure}[H]
\centering
\includegraphics{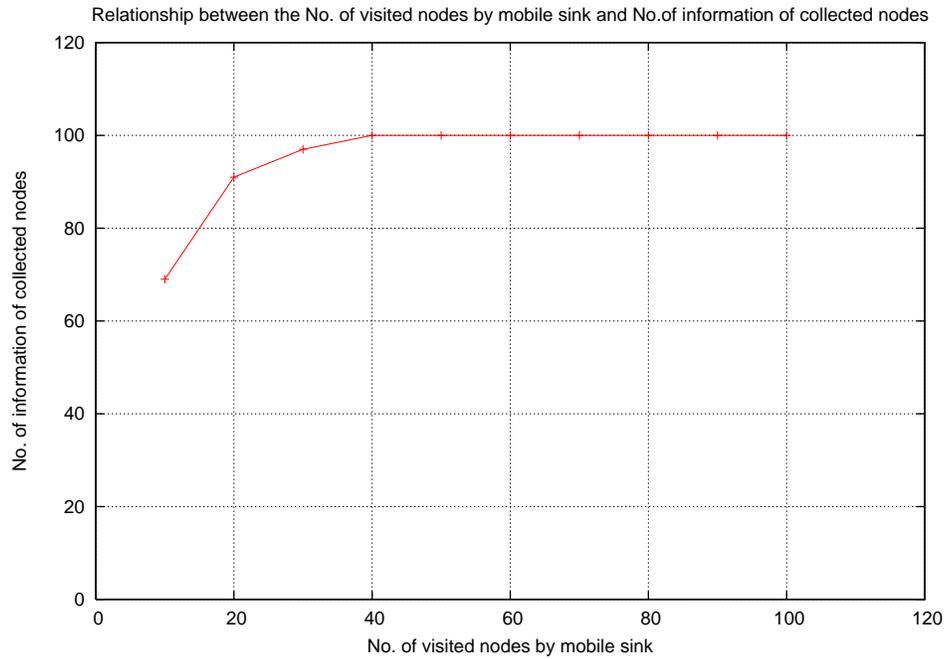}
\vspace{40pt}
\caption{Relationship between the No. of visited nodes by mobile sink and No. of information of collected nodes when all nodes are active}
\label{rwallactive}
\end{figure}

Here in Fig. ~\ref{rwallactive}, we make all the nodes active, that is, we did not consider active sleep regieme and we found that the results are quite interesting. By visiting only a very few number of nodes, mobile sink is able to get approximately 80\% of the representative view of the network.\\

\begin{figure}[H]
\centering
\includegraphics{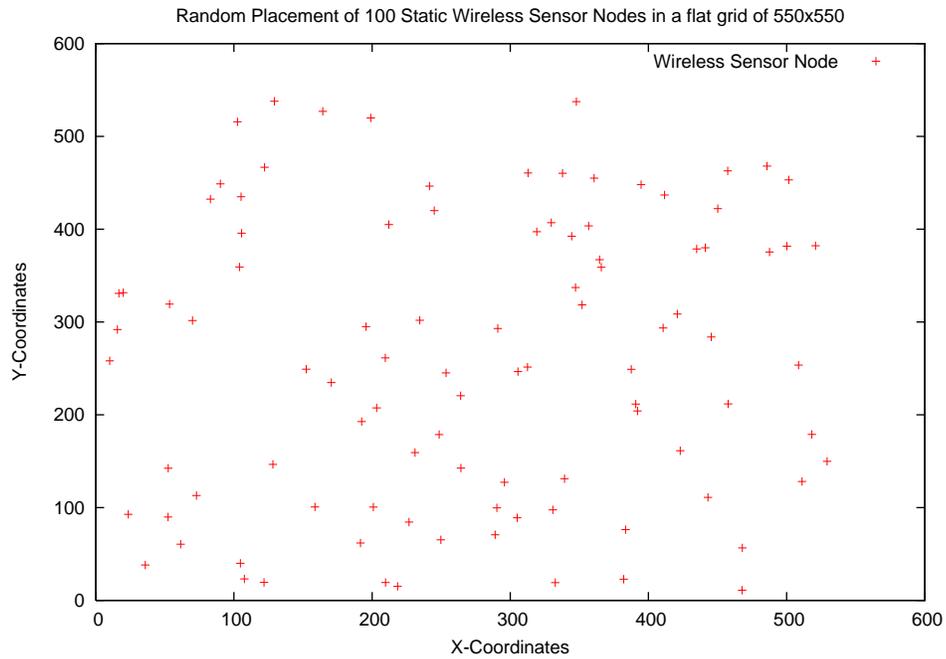}
\vspace{40pt}
\caption{Static Wireless Sensor Nodes placement on a flat grid of 550x550}
\label{rwdesnity}
\end{figure}

\pagebreak
Here in this Fig. ~\ref{rwdesnity}, we are combining three parameters, small grid size of 550x550, very small time out between 1 and 2, and very high node degree which is about 44 and for that sake we have changed the topology and we randomly placed static wireless sensor nodes on the flat grid.

\begin{figure}[H]
\centering
\includegraphics{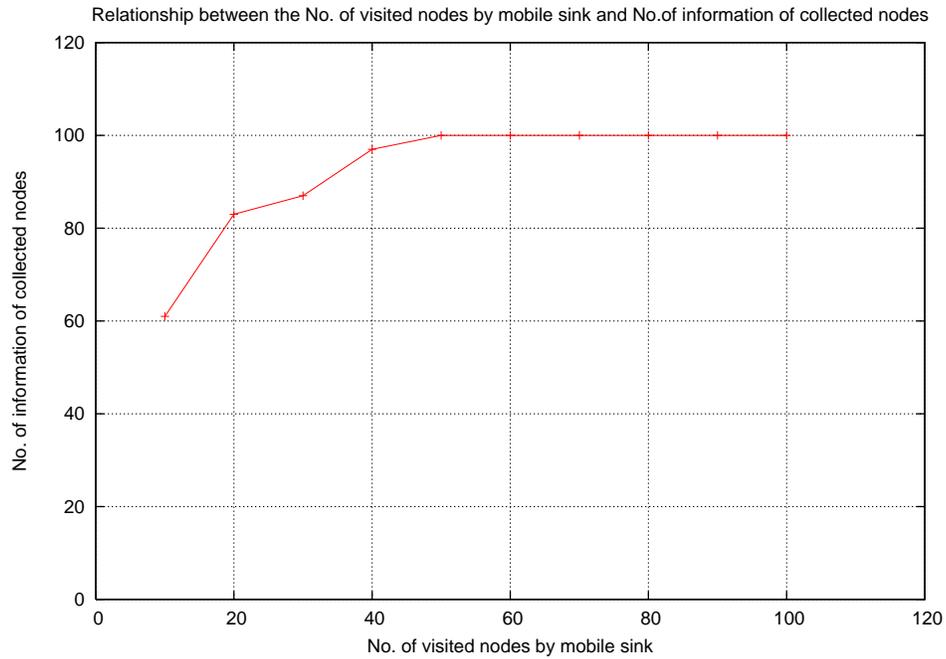}
\vspace{40pt}
\caption{Impact of small grid size, very small time out, and very high node degree}
\label{rwdgraph}
\end{figure}

In this Fig. ~\ref{rwdgraph}, we are trying to vary three parameters and trying to find the trade-off between them and found that the results are interesting. We found that all the three factors plays an important role in collection of data by mobile sink. So, we are analyzing these parameters and the work is still going on, but due to lack of time, we were just able to present here few results.

\pagebreak

\subsection{Factors influencing number of information of collected nodes by mobile sink}
We observed that number of information of collected nodes from the network by the mobile sink depends upon the following factors \footnote{We are planning to analyze the impact of these factors and the work is going on !}:
\begin{itemize}

\item View size: We are assuming two types of views, namely \emph{time based} and \emph{size based}. In \emph{time based}, the node discards all views from its view that the node has not heard from the last timeout period. While in \emph{size based} strategy, a node maintains a restriction on its view by counting the number of views it can keep in its memory.
\item Intersection of views: While performing RW, a node may end their RW on two different nodes, which results in the intersection of views. As the goal of our proposal, data from a same node will be disseminated by RWs in different storage nodes in the network. Although an important property for the sucess of our proposal, a high level of intersection of views of different nodes may impact the
data collection by the sink. In particular, a high percentage of common information at nodes, reduces the number of collected information by visited nodes.
\item Convergence time: It is the time required until all views reach their target size, which can vary and depends upon \emph{time based} or \emph{size based} strategy of maintaining views. 
\item Node density: Here it means that averagely how many nodes are neighbours of a particular node. If there are less neighbours then there will be definitely a impact on the RW.
\item Active/Sleep Regime (Value of $\delta$): It means that for how much time a sensor node will active and switch to sleep state periodically. A small activity period will slow down the data distribution in the network, resulting in empty views, and by consequence, few information to be collected.
\item Neighbour discovery through Hello Packets (impact of receive/send Hello packets in active state): Neighbours can be discovered only when the nodes are active and turn on their radio and depends upon the interval of sending of Hello packets. A problem in neighborhood discovery will impact the RW performance.
\item Impact of timeout state of nodes: Timeout state of nodes means that the time when node initialize. 
\item Random walk length: It can be n, n/2, n/4 etc. and it means that each node will take how many steps to reach a random node. Shorter RW will not allow the uniformely membership distribution as specified by RaWMS.
\item Time a mobile sink visits (sojourn time): It is a time that mobile sinks spends in traversing and collecting the information from the network.
A shorter period of visits will result in few data collection.

\end{itemize}
\pagebreak

\section{Discussion and Open Issues}
\label{sec:discussion}

\subsection{Discussion}

Table~\ref{table_example1} represents a comparison of different Proactive Data-Dissemination
Approaches. In Table~\ref{table_example1}, we discuss Hotspots problem for each protocol. Many
proposed protocols uses cluster based approach or suggested that some nodes should be responsible
for forwarding data of all other nodes, which results in Hotspots problem.

We then describe multi-hop routing. Here, multi-hop routing means whether nodes needs to
construct a path by routing packets through the network to other nodes or to the mobile sink.
Storage Motes parameter indicates the use of particular nodes to store data in the network. We also
mentioned Active/Sleep States of Wireless Sensor Nodes  because this has a great impact on the
overall lifetime of the network.  The more the sensor nodes are active, the less the network
lifetime.


\begin{table}[H]
\renewcommand{\arraystretch}{2}

\caption{Proactive Approaches}

\label{table_example1}
\centering
\begin{tabular}{p{3.2cm}||p{1.3cm}||p{1.6cm}||p{1.0cm}||p{1.7cm}||p{1.7cm}}

\hline

\bfseries \footnotesize Protocol Name & \bfseries \footnotesize Hot Spot Problem & \bfseries \footnotesize Multi-hop Routing & \bfseries \footnotesize Storage Motes & \bfseries \footnotesize Active Sleep State of Sensor Nodes & \bfseries \footnotesize Active Sleep State of Storage Motes \\
\hline\hline

\footnotesize Our Proposal & \footnotesize No & \footnotesize No & \footnotesize Yes & \footnotesize Yes & \footnotesize Yes\\
\hline
\footnotesize Moving Schemes for Mobile Sink in WSN & \footnotesize No & \footnotesize Yes & \footnotesize No & \footnotesize Didn't mention & \footnotesize Didn't mention\\
\hline
\footnotesize Coordinate Magnetic Routing for MSWSN & \footnotesize Yes & \footnotesize Yes & \footnotesize No & \footnotesize No & \footnotesize No\\
\hline
\footnotesize FLOW  & \footnotesize Yes & \footnotesize Yes & \footnotesize No & \footnotesize No & \footnotesize No\\
\hline
\footnotesize WEDAS & \footnotesize Yes & \footnotesize Yes & \footnotesize No & \footnotesize No & \footnotesize No\\

\hline

\end{tabular}
\end{table}



\begin{table}[H]
\renewcommand{\arraystretch}{2}

\caption{Proactive Approaches showing features of Mobile Sink}

\label{table_example2}
\centering
\begin{tabular}{p{3.0cm}||p{1.2cm}||p{2.0cm}||p{1.8cm}||p{3.2cm}}

\hline

\bfseries \footnotesize Protocol Name & \bfseries \footnotesize Multiple Mobile Sink & \bfseries \footnotesize Mobile Sink Trajectory & \bfseries \footnotesize Sink Traverse the Whole Network & \bfseries \footnotesize Sink Broadcast Messages to collect information like energy consumption \\
\hline\hline

\footnotesize Our Proposal& \footnotesize No & \footnotesize Free to Move  & \footnotesize No & \footnotesize No Broadcasting is needed\\
\hline
\footnotesize Moving Schemes for Mobile Sink in WSN & \footnotesize No & \footnotesize Based on sensor node position and energy level  & \footnotesize No & \footnotesize Sink broadcasts its location and in reply sensor nodes send their position and energy level\\
\hline
\footnotesize Coordinate Magnetic Routing for MSWSN & \footnotesize No  & \footnotesize Trajectory depends upon received data & \footnotesize No & \footnotesize Sink have to broadcast HELLO Messages to neighboring CHs to notify its location\\
\hline
\footnotesize FLOW & \footnotesize No & \footnotesize Fixed Trajectory & \footnotesize No just predefined path in a periodic way & \footnotesize Mobile Sink will not query any data. It is the sensors who when sensed data, calulate the route and then send data to the Mobile Sink\\
\hline
\footnotesize WEDAS & \footnotesize No & \footnotesize RWP Model. They did not provide any improvement & \footnotesize Yes & \footnotesize Sink advertises its current and future position to its neighbors.\\

\hline

\end{tabular}
\end{table}


In Table~\ref{table_example2}, we are presenting a summary of the proactive approaches of data
dissemination schemes keeping in mind the features of mobile sink. As you can see we firstly
mentioned whether the protocol needs a single mobile sink or multiple mobile sinks to accomplish
its task. Secondly, we also indicated that type of mobile sink's trajectory. Thirdly, we mentioned
whether the mobile sink will traverse the whole network or just follow a certain trajectory to
collect data. Finally, we mentioned the broadcasting mechanism followed by sink.

We observe that except WEDAS \cite{IEEEhowto:habib}, all the other studied proactive approaches
does not require the sink to traverse the whole network, which is an optimization in terms of
mobile sink trajectory.



\begin{table}[H]
\renewcommand{\arraystretch}{2}

\caption{Reactive Approaches}

\label{table_example3}
\centering
\begin{tabular}{p{3.2cm}||p{1.3cm}||p{1.6cm}||p{1.0cm}||p{1.7cm}||p{1.7cm}}
\hline

\bfseries \footnotesize Protocol Name & \bfseries \footnotesize Hot Spot Problem & \bfseries \footnotesize Multi-hop Routing & \bfseries \footnotesize Storage Motes & \bfseries \footnotesize Active Sleep State of Sensor Nodes & \bfseries \footnotesize Active Sleep State of Storage Motes \\
\hline\hline

\footnotesize Locators of Mobile Sink for WSNs & \footnotesize Yes & \footnotesize Yes & \footnotesize No & \footnotesize No & \footnotesize No\\
\hline
\footnotesize Interest Dissemination with Directional Antennas for WSNs with Mobile Sinks & \footnotesize Yes & \footnotesize Yes & \footnotesize No & \footnotesize Yes & \footnotesize No\\
\hline
\footnotesize Data MULES & \footnotesize No & \footnotesize No & \footnotesize Yes & \footnotesize Did not mention & \footnotesize Did not mention\\
\hline
\footnotesize MobiRoute  & \footnotesize Yes & \footnotesize Yes & \footnotesize Yes & \footnotesize Did not mention & \footnotesize Did not mention\\
\hline
\footnotesize Efficient Data Propagation Strategies in WSNs using a Mobile Sink & \footnotesize Yes & \footnotesize Yes & \footnotesize No & \footnotesize No & \footnotesize No\\

\hline

\end{tabular}
\end{table}


In the same way, Table \ref{table_example3} represents comparison of reactive data dissemination
approaches and Table \ref{table_example4} summarizes the features of data dissemination schemes
keeping in mind the features of mobile sink.

\begin{table}[H]
\renewcommand{\arraystretch}{2}

\caption{Reactive Approaches showing features of Mobile Sink}

\label{table_example4}
\centering
\begin{tabular}{p{3.0cm}||p{1.2cm}||p{2.0cm}||p{1.8cm}||p{3.2cm}}

\hline

\bfseries \footnotesize Protocol Name & \bfseries \footnotesize Multiple Mobile Sink & \bfseries \footnotesize Mobile Sink Trajectory & \bfseries \footnotesize Sink Traverse the Whole Network & \bfseries \footnotesize Sink Broadcast Messages to collect information like energy consumption \\
\hline\hline

\footnotesize Locators of Mobile Sink for WSNs & \footnotesize Yes & \footnotesize Did not talk about Mobile Sink Trajectory & \footnotesize No & \footnotesize Broadcast messages to propagate sink's location. \\
\hline
\footnotesize Interest Dissemination with Directional Antennas for WSNs with Mobile Sinks & \footnotesize No & \footnotesize Sink has to follow a particular trajectory & \footnotesize Yes & \footnotesize Interest packets are broadcasted periodically using a directional antenna in order to set up routes in the network before sinks arrival.\\
\hline
\footnotesize Data MULES & \footnotesize Considered both cases of Single and multiple sinks & \footnotesize Random Walk Mobility & \footnotesize Yes & \footnotesize Did not mention. \\
\hline
\footnotesize MobiRoute & \footnotesize No & \footnotesize No solution provided for sink mobility, instead provided support for sink mobility in terms of link breakage and less packet loss & \footnotesize Yes & \footnotesize Yes\\
\hline
\footnotesize Efficient Data Propagation Strategies in WSNs using a Mobile Sink & \footnotesize No & \footnotesize Improved not only sink trajectory but also optimize data aggregation techniques & \footnotesize No & \footnotesize No\\

\hline

\end{tabular}
\end{table}



\begin{table}[H]
\renewcommand{\arraystretch}{2}

\caption{Optimization of Mobile Sink Trajectory and Data-Dissemination}

\label{table_example5}
\centering
\begin{tabular}{p{5cm}||p{2.4cm}||p{2.8cm}}

\hline

\bfseries \footnotesize {Protocol Name} & \bfseries  \footnotesize Optimize Mobile Sink Trajectory & \bfseries \footnotesize Optimize Data-Dissemination Technique \\
\hline\hline

\footnotesize Our Proposal & \footnotesize No & \footnotesize Yes\\
\hline
\footnotesize Moving Schemes for Mobile Sink in WSN & \footnotesize Yes & \footnotesize No\\
\hline
\footnotesize Coordinate Magnetic Routing for MSWSN & \footnotesize No & \footnotesize Yes\\
\hline
\footnotesize FLOW  & \footnotesize No & \footnotesize Yes\\
\hline
\footnotesize WEDAS & \footnotesize No & \footnotesize Yes\\
\hline
\footnotesize Locators of Mobile Sink for WSNs & \footnotesize No & \footnotesize Yes \\
\hline
\footnotesize Interest Dissemination with Directional Antennas for WSNs with Mobile Sinks & \footnotesize No & \footnotesize Yes\\
\hline
\footnotesize Data MULES & \footnotesize No & \footnotesize No\\
\hline
\footnotesize MobiRoute  & \footnotesize No & \footnotesize No\\
\hline
\footnotesize Efficient Data Propagation Strategies in WSNs using a Mobile Sink & \footnotesize Yes & \footnotesize Yes\\
\hline

\end{tabular}
\end{table}


Table \ref{table_example5} represents which protocol optimizes mobile sink trajectory and which
protocol optimizes data-dissemination technique.

\subsection{Open Issues}
The incorporation of Mobile Sink in Wireless Sensor Network for Data Collection has introduced many
challenges. Although a lot of work has been done, data-dissemination and data-management with
mobile sink is in developing stages. In this section, we are discussing open issues that are remain
to be addressed.

\begin{itemize}
\item \emph{Issue 1} : Distributed Data Storage Capability

Distributed data storage capability plays an important role in the presence of Mobile Sink. In
particular how to safely store collected data such that they can be retrieved later. Although we
have given a solution to a certain extent to deal with this issue, there is still a need to further
investigate this issue by incorporating other parameters like node density, speed of mobile sink,
etc.



\item \emph{Issue 2} : Speed of Mobile Sink

There is also an influence of speed of Mobile Sink on data collection. If the speed of Mobile Sink
is so much high then definitely there will be a high packet loss. So, there is also a need to
investigate this issue.

\item \emph{Issue 3} : Sink Traversal Time

Traversing the network in a timely and in an efficient way of Mobile Sink is critical since failure
to visit  some areas of the network will result in data loss while if there is a gap in mobile
sinks visit for collection of data then it will result in receiving of non updated data.

\item \emph{Issue 4} : Multiple Sinks

It is evident that those schemes that just only support single Mobile Sink will not work
efficiently in the  presence of multiple mobile sinks. So investigating the impact of multiple
mobile sinks is an open research issue that need to be addressed.

\item \emph{Issue 5} : Security

There are applications where data integrity and security is crucial like behind the enemy lines or
inhospitable  terrains. So, security is an hot issue that need to addressed.

\item \emph{Issue 6} : Nodes Density

There is a need to investigate the impact of nodes density on data collection by mobile sink
because nodes density  impacts the dissemination mechanism in terms of convergence or overhead.
\end{itemize}



%
%

\section{Conclusion}
\label{sec:conclusion} Data Dissemination with Mobile Sinks in Wireless Sensor Networks has
attracted many researchers because incorporation  of mobile sink minimizes the number of
transmissions, eliminating the redundant data, conserve the energy, and thus resulting in the
overall increase in the lifetime of the network.

In this report, firstly, we presented state of the art survey on Data Management and Data
Dissemination techniques  with Mobile Sink. Moreover we classified these techniques into two ample
sub-categories. Under this classification, we identify, review, compare, and highlight these
techniques and their pros and cons. We did a SWOT (Strength, Weaknesses, Opportunities, Threats)
analysis of each scheme. We also discussed where each scheme is appropriate.

Secondly we presented a new \emph{distributed data management} scheme which is based upon Random
Walk Based Membership  Service to allow Data Dissemination in Mobile Sink based Wireless Sensor
Networks. Our proposed scheme efficiently deals with the aforementioned problems that we discussed
in this report and we also discuss the characteristics of our proposed scheme compared with the
state of the art data dissemination schemes.

To the best of our knowledge, we are the first to propose an efficient data dissemination approach
(in terms of overhead, adaptiveness and representativeness) to allow a mobile sink to gather a
representative view of the monitored region covered by $n$ sensor nodes by visiting any $\sqrt{n}$
nodes. In addition, our proposed mechanism also allows node to switch between active and sleep
state and still guarantee the minimum required $\sqrt{n}$ active nodes in the network.

We have analyzed the behavior of this active/sleep regime, where the average number of active
nodes at discrete points of simulation time as a function of an increasing $\delta$ (from 0 to 1)
were extracted. The objective was to determine the upper bound on the $\delta$ fraction of nodes
that may be put to sleep, in order to guarantee at least $\sqrt{n}$ nodes awaken in the network at
any given time.

Once the $\delta$ parameter was well analyzed, the data was disseminated in the network by
generating random walks for each node using RW distance of $d = \frac{n}{2}$ and view size of $k =
\sqrt{n}$. The behavior of the data dissemination using random walks was then analyzed. This has
consisted in making sink to visit $\sqrt{n}$ nodes and verifying the information of how many nodes
were really collected.

After numerious simulations, we found that a mobile sink is able to collect the 
representative view of the monitored region covered by $n$ sensor nodes by only visiting {\it any} $m$ nodes, where $m << n$. 



%
%

%









\bibliographystyle{IEEEtran} 
\bibliography{mubashirbib} 

\begin{thebibliography}{10}
\providecommand{\url}[1]{#1}
\csname url@samestyle\endcsname
\providecommand{\newblock}{\relax}
\providecommand{\bibinfo}[2]{#2}
\providecommand{\BIBentrySTDinterwordspacing}{\spaceskip=0pt\relax}
\providecommand{\BIBentryALTinterwordstretchfactor}{4}
\providecommand{\BIBentryALTinterwordspacing}{\spaceskip=\fontdimen2\font plus
\BIBentryALTinterwordstretchfactor\fontdimen3\font minus
  \fontdimen4\font\relax}
\providecommand{\BIBforeignlanguage}[2]{{%
\expandafter\ifx\csname l@#1\endcsname\relax
\typeout{** WARNING: IEEEtran.bst: No hyphenation pattern has been}%
\typeout{** loaded for the language `#1'. Using the pattern for}%
\typeout{** the default language instead.}%
\else
\language=\csname l@#1\endcsname
\fi
#2}}
\providecommand{\BIBdecl}{\relax}
\BIBdecl

\bibitem{IEEEhowto:yanzhong}
Y.~Bi, J.~Niu, L.~Sun, W.~Huangfu, and Y.~Sun, ``Moving schemes for mobile
  sinks in wireless sensor networks,'' in \emph{Performance, Computing, and
  Communications Conference}.\hskip 1em plus 0.5em minus 0.4em\relax IPCCC
  2007, IEEE, 2007.

\bibitem{IEEEhowto:shih}
S.~H. Chang, M.~Merabti, and H.~M. Mokhtar, ``Coordinate magnetic routing for
  mobile sink wireless sensor networks,'' in \emph{21st International
  Conference on Advance Information Neworking and Applications Workshops (AINAW
  07)}, 2007 IEEE.

\bibitem{IEEEhowto:rahul}
R.~Urgaonkar and B.~Krishnamachari, ``Flow: An efficient forwarding scheme to
  mobile sink in wireless sensor networks,'' in \emph{First IEEE International
  Conference on Sensor and Ad hoc Communications and Networks (SECON 2004)},
  Santa Clara, CA, October 2004.

\bibitem{IEEEhowto:habib}
H.~M. Ammari and S.~K. Das, ``Data dissemination to mobile sinks in wireless
  sensor networks : An information theoretic approach (wedas),'' in \emph{MASS
  2005}, 2005 IEEE.

\bibitem{IEEEhowto:guydong}
G.~Shim and D.~Park, ``Locators of mobile sink for wireless sensor networks,''
  in \emph{Proceedings of the 2006 International Conference on Parallel
  Processing Workshops (ICPPW06)}, 2006 IEEE.

\bibitem{IEEEhowto:yihong}
Y.~Wu, L.~Zhang, Y.~Wu, and Z.~Niu, ``Interest dissemination with directional
  antennas for wsns with mobile sinks,'' in \emph{SenSys06}, Boulder, Colorado,
  USA, November 1-3, 2006.

\bibitem{IEEEhowto:rc}
R.~C. Shah, S.~Roy, S.~Jain, and W.~Brunette, ``Data mules: Modeling and
  analysis of a three-tier architecture for sparse sensor networks,''
  \emph{Elsevier Ad Hoc Networks Journal}, vol. 1,, issues 2-3, pp. 215--233,
  Sept. 2003.

\bibitem{IEEEhowto:jun}
J.~Luo, J.~Panchard, M.~Piorkowski, M.~Grossglauser, and J.~P. Hubaux,
  ``Mobiroute : Routing towards a mobile sink for improving lifetime in sensor
  networks,'' in \emph{Proceedings of the 2nd IEEE/ACM International Conference
  on Distributed Computing in Sensor Systems(DCOSS06)}, San Francisco,
  California, USA, June 2006.

\bibitem{IEEEhowto:ioannis}
I.Chatzigiannakis, A.Kinalis, and S.Nikoletseas, ``Efficient data propagation
  strategies in wireless sensor networks using a single mobile sink,'' \emph{In
  the Journal of Computer Communications (COMCOM), Elsevier}, vol. 31 (5), pp.
  896--914, 2008.

\bibitem{IEEEhowto:anis}
A.~A. et~al, ``A line in the sand: A wireless sensor network for target
  detection, classification, and tracking,'' \emph{The International Journal of
  Computer and Telecommunications, Special issue: Military communications
  systems and technologies}, vol. 46, Issue 5, pp. 605--634, 2004.

\bibitem{IEEEhowto:gyula}
G.~S. et~al, ``Sensor network based counter sniper system,'' in
  \emph{Proceedings of the 2nd international conference on Embedded networked
  sensor systems}, Baltimore, MD, USA, 2004.

\bibitem{IEEEhowto:i}
I.~Akyildiz, T.~Melodia, and K.~R. Chowdhury, ``A survey on wireless multimedia
  sensor network,'' \emph{The International Journal of Computer and
  Telecommunications Networking}, vol. 51, Issue 4, pp. 921--960, 2007.

\bibitem{IEEEhowto:david}
D.~Culler, D.~Estrin, and M.~Srivastava, ``Guest editors' introduction:
  Overview of sensor networks,'' \emph{IEEE Computer}, vol. 37, No 8, August
  2004.

\bibitem{IEEEhowto:bg}
L.~Barichello, R.~Garcia, C.~Siewert, and R.~W. T.~Hill, ``Swot analysis: Its
  time to product recall,'' \emph{Long Range Planning, Elsivier}, vol. 30,
  Number 1, pp. 4--5+46(2), 1997.

\bibitem{IEEEhowto:aka}
A.~Kansal, A.~Somasundara, D.~Jea, M.~Srivastava, and D.~Estrin, ``Intelligent
  fluid infrastructure for embedded networking,'' in \emph{ACM MobiSys}, June
  2004.

\bibitem{IEEEhowto:zbar}
Z.~B. Yossef, R.~Friedman, and G.~Kliot, ``Rawms random walk based lightweight
  membership service for wireless ad hoc networks,'' in \emph{7th ACM
  International Symposium on Mobile Ad Hoc Networking and Computing(MobiHoc)},
  Florence, Italy, May 2006.

\bibitem{IEEEhowto:aw}
A.~Woo, T.~Tong, and D.~Culler, ``Taming the underlying challenges of reliable
  multihop routing in sensor networks,'' in \emph{1st ACM SenSys}, 2003.

\bibitem{IEEEhowto:jens}
J.~B. Schmitt, F.~A. Zdarsky, and utz Roedig, ``Sensor network calculus with
  multiple sinks,'' in \emph{Proceedings of IFIP NETWORKING 2006, Workshop on
  Performance Control in Wireless Sensor Networks}.\hskip 1em plus 0.5em minus
  0.4em\relax Coimbra, Portugal: Springer LNCS, May 2006, p. 613.

\end{thebibliography}

\pagebreak
\pagestyle{empty}
 
\section{Appendex}

\algrenewcommand{\algorithmiccomment}[1]{ \hfill $\rightarrow$ #1

}

\noindent\begin{minipage}[b]{\linewidth}
 \begin{algorithm}[H]
  \caption{Code for a Sensor Node}\label{sensorAlg}
 \begin{algorithmic}[1]

\footnotesize
\State \textbf{do} forever
\State \hspace{0.5in}	wait (\emph{U} time units)
\State \hspace{0.5in}	//start a new RW
\State \hspace{0.5in}	ttl $\leftarrow MixingTime;$
\State \hspace{0.5in}	$viewsize \leftarrow \sqrt{n};$
\State \hspace{0.5in}	HandleRW(myaddr,ttl,current\_time);
\State \textbf{end do}\\

\State \textbf{do} forever

\State \hspace{0.5in}	//send Hello Packets for neighbour discovery

\State \hspace{0.5in}	\textbf{if}(myaddrs.status==ACTIVE)

\State \hspace{1.0in}		sendHello();

\State \hspace{0.5in}	\textbf{else}

\State \hspace{1.0in}		do nothing;

\State \textbf{end do}\\

\State HandleRW(myaddr,ttl,current\_time)

\State \hspace{0.5in}	\textbf{if}(ttl==0)

\State \hspace{1.0in}	//node finished its RW
	
\State \hspace{0.5in}	\textbf{while}(ttl\textgreater0) do

\State \hspace{1.0in}		$next \leftarrow PickNextNode(myaddr);	$

\State \hspace{1.0in}		$hasView \leftarrow next;$

\State \hspace{1.0in}		\textbf{if}(next != myaddr)

\State \hspace{1.5in}			send(RWmessage\textless myaddr,ttl,current\_time\textgreater) to next;

\State \hspace{1.5in}		$	myaddr \leftarrow next;$

\State \hspace{1.0in}		\textbf{else}

\State \hspace{1.5in}			ttl=ttl-1;

\State \hspace{1.0in}		\textbf{end if	}

\State \hspace{0.5in}	//publishes view \textless myaddr, DATA, current\_time\textgreater

\State \hspace{0.5in}	publishView(hasView); \\

\State publishView(hasView)

\State \hspace{0.5in} //size based method
	
\State \hspace{0.5in}	\textbf{if}\emph{sizeBasedMethod} that is (viewsize\textgreater $\sqrt{n}$) then 

\State \hspace{1.0in}		discardOldestFromView(hasView);

\State \hspace{0.5in} //timeout based method
	
\State \hspace{0.5in}	\textbf{if} \emph{timeoutBasedMethod} then 

\State \hspace{1.0in}		discardExpiredFromView(hasView,timeout);\\

\State PickNextNode(myaddr)

\State \hspace{0.5in}	\textbf{if}(myaddr.neighbourlist.length==0)

\State \hspace{1.0in}		return myaddr;

\State \hspace{0.5in}	\textbf{else if}(myaddr.neighbourlist.length==1)

\State \hspace{1.0in}		$neighbourid \leftarrow get\_neighbour$
		
\State \hspace{1.5in}		\textbf{if}(neighbourid.status==TIMEOUT)

\State \hspace{2.0in}			return myaddr;

\State \hspace{1.5in}		\textbf{else if}(neighbourid.status==SLEEP)

\State \hspace{2.0in}			return myaddr;

\State \hspace{1.5in}		\textbf{else if}(neighbourid.status==ACTIVE)

\State \hspace{2.0in}			return neighbourid;

\State \hspace{1.5in}		\textbf{end if}

\State \hspace{0.5in}	\textbf{else if}(myaddr.neighbourlist.length\textgreater 1)

\State \hspace{1.0in}		//select any random neighbour and check its status

\State \hspace{1.0in}		$r\_neighbourid \leftarrow get\_randomly\_selected\_neighbour$
		
\State \hspace{1.5in}		\textbf{if}(r\_neighbourid.status==TIMEOUT)

\State \hspace{2.0in}			return myaddr;

\State \hspace{1.5in}		\textbf{else if}(r\_neighbourid.status==SLEEP)

\State \hspace{2.0in}			return myaddr;

\State \hspace{1.5in}		\textbf{else if}(r\_neighbourid.status==ACTIVE)

\State \hspace{2.0in}			return r\_neighbourid;

\State \hspace{0.5in}	\textbf{end if}

\end{algorithmic}
\end{algorithm}
\end{minipage}

\end{document}